\documentstyle[colordvi]{mn}    
\def\kms{$\rm km\;s^{-1}$}    
\def\dg{^\circ}    
\def\ha{H$\alpha$}    
\def\h2{H$_{2}$}    
\def\hi{H\,{\scriptsize I}}    
\def\hii{H\,{\scriptsize II}}    
\def\nii{[N\,{\scriptsize II}]}    
\def\oiii{[O\,{\scriptsize III}]$\,\lambda5006.8$}    
    
\def\msun{M$_{\odot}$}    
\def\Msun{M$_{\odot}$}  
\def\lvsun{L$_{\odot,V}$}    
    
\def\mas{mag arcsec$^{-2}$}    
    
\begin{document}    
    
    
\title[NGC 772, NGC 3898, and NGC 7782]
{Modelling gaseous and stellar kinematics in the disc galaxies NGC
772, NGC 3898, and NGC 7782\thanks{Based on observations carried out
at the Multiple Mirror Telescope Observatory (Arizona, USA) which is a
joint facility of the Smithsonian Institution and the University of
Arizona, with the Isaac Newton Telescope which is operated on the
island of La Palma by the Isaac Newton Group in the Spanish
Observatorio del Roque de los Muchachos of the Instituto de Astrof\'\i
sica de Canarias, and with the Vatican Advanced Technology Telescope,
which is the Alice P. Lennon Telescope and the Thomas J. Bannan
Astrophysics Facility at the Mount Graham International Observatory
(Arizona, USA).}
$^{\bf,}$ \thanks{Tables 4 to 12 are only available in    
electronic form at the CDS via anonymous ftp to  
cdsarc.u-strasbg.fr (130.79.128.5) or via   
http://cdsweb.u-strasbg.fr/Abstract.html.}
}

    
\author[E. Pignatelli et al.]{    
     E. Pignatelli$^1$\thanks{email:{\tt pignatel@sissa.it}},    
     E.M.~Corsini$^2$,    
     J.C. Vega Beltr\'an$^3$,  
     C.~Scarlata$^4$,  
     A.~Pizzella$^2$,         
     \newauthor   
     J.G.~Funes, S.J.$^5$,         
     W.W. Zeilinger$^6$,    
     J.E. Beckman$^3$ 
     and F.~Bertola$^4$\\  
     $^1$SISSA, via Beirut 2-4, I-34013 Trieste, Italy \\   
     $^2$Osservatorio Astrofisico di Asiago, Dipartimento di Astronomia,    
         Universit\`a di Padova,     
         via dell'Osservatorio~8, I-36012 Asiago, Italy \\    
     $^3$Instituto Astrof\'\i sico de Canarias, Calle Via Lactea s/n,    
         E-38200 La Laguna, Spain \\    
     $^4$Dipartimento di Astronomia, Universit\`a di Padova,     
         vicolo dell'Osservatorio~5, I-35122 Padova, Italy \\   
     $^5$Vatican Observatory, University of Arizona,     
         Tucson, AZ 85721, USA \\   
     $^6$Institut f{\"u}r Astronomie, Universit{\"a}t Wien,    
         T{\"u}rkenschanzstra{\ss}e 17, A-1180 Wien, Austria  
}


\date{Received..................; accepted...................}    
    
    
\maketitle    
   
\begin{abstract}    
 
We present $V-$band surface photometry and major-axis kinematics of
stars and ionized gas of three early-type spiral galaxies, namely NGC
772, NGC 3898 and NGC 7782.  For each galaxy we present a
self-consistent Jeans model for the stellar kinematics, adopting the
light distribution of bulge and disc derived by means of a
two-dimensional parametric photometric decomposition.  This allowed us
to investigate the presence of non-circular gas motions, and derive
the mass distribution of luminous and dark matter in these objects.

NGC 772 and NGC 7782 have apparently normal kinematics with the
ionized gas tracing the gravitational equilibrium circular speed.
This is not true in the innermost region ($|r|\la8''$) of NGC 3898
where the ionized gas is rotating more slowly than the circular
velocity predicted by dynamical modelling.  This phenomenon is common
in the bulge-dominated galaxies for which dynamical modelling enables
us to make the direct comparison between the gas velocity and the
circular speed, and it poses questions about the reliability of galaxy
mass distributions derived by the direct decomposition of the observed
ionized-gas rotation curve into the contributions of luminous and dark
matter.

\end{abstract}     
    
    
\section{Introduction}    
\label{sec:introduction}    
    
In the past kinematic and dynamical studies of disc
galaxies have been mainly focused on late-type
spirals. With respect to lenticular and early-type spiral
galaxies (defined as those with $B/T>0.3$, following
Simien \& de Vaucouleurs 1986), the study of late-type
spirals has many advantages from both the observational
and interpretative points of view. Late-type spirals have
small or even negligible bulges, so they are usually
described as luminous discs embedded in dark matter
halos.  They are on average more gas-rich than early-type
spirals. Gas is mainly confined in the galactic disc, 
considered moving at circular speed and therefore used
as direct tracer of the gravitational potential. Since
the line-of-sight velocity distribution (LOSVD hereafter)
of ionized or neutral hydrogen is easier to measure
than the stellar one, the number of spirals of known
gaseous kinematics increased faster than that of the
galaxies for which stellar motions were measured.  These
facts made late-type spirals easy targets, particularly in
the search for dark matter (see Sofue \& Rubin 2001 for a
review).

Only recently Heraudeau \& Simien (1998) and Heraudeau et
al. (1999) measured the stellar rotation curves and
velocity-dispersion profiles along the major axis of a large sample of
spirals (more than 60 objects) ranging from Sa to Sc.  In last years
systematic surveys have been started to derive the stellar and gaseous
kinematics in S0's (Bertola et al. 1995; Fisher 1997; Bettoni \&
Galletta 1997) and in early-to-intermediate type spirals (Corsini et
al. 1999; Vega Beltr\'an 1999; Vega Beltr\'an et al. 2000).
 
If the direct comparison of stellar and gaseous kinematics allowed to 
unveil also in disc galaxies the presence of kinematically-decoupled 
components (see Bertola \& Corsini 1999 for a recent review), the 
application of dynamical models showed that in the bulge-dominated 
region the ionized-gas velocity may fall below the circular speed 
(Fillmore, Boroson \& Dressler 1986; Kent 1988; Kormendy \& Westpfahl 
1989). These `slowly-rising' rotation curves have 
been interpreted as the signature of the 
presence of pressure-supported gas (Bertola et al. 1995; Cinzano et 
al. 1999), which has been recently expelled by stars of the bulge 
but not yet heated to the virial temperature of the galaxy. This 
discovery poses new questions about the reliability of galaxy mass 
distributions derived by the decomposition of the 
observed ionized-gas rotation curve into the contributions of
luminous and dark matter. In fact the inner 
gradient of the gas velocity curve is usually used to fix the amount 
of luminous matter, which results to be underestimated if gas velocity 
rises slowly than circular speed. 
 
In this paper we present a study of the ionized-gas and stellar
kinematics for 3 Sa -- Sb galaxies, namely NGC 772, NGC 3898, and NGC
7782. We apply a self-consistent dynamical model based on a 
two-dimensional photometric decomposition method to explain the different
kinematical behaviour of gas and stars.  The goal of this work is to
derive the mass distribution of these objects and investigate the
possible presence of non-circular gas motions in the bulge region.
 
The paper is organized as it follows. In Sect.~\ref{sec:properties} we 
give an overview of the properties of NGC 772, NGC 3898 and NGC 7782; 
in Sect.~\ref{sec:observations} we present the broad and narrow-band 
imaging and major-axis kinematics of stars and ionized gas; in 
Sect.~\ref{sec:structure} we discuss the photometric and dynamical 
techniques adopted to investigate the light and mass distribution of 
these galaxies; in Sect.~\ref{sec:results} we apply these techniques 
to the galaxies, deriving their structural and kinematic parameters; 
finally, Sect.~\ref{sec:conclusions} is devoted to the discussion of 
the results and conclusions. 
 
\section{Global properties of NGC 772, NGC 3898 and NGC 7782}  
\label{sec:properties}    
    
The galaxies studied in this paper are a subset of the 20 disc 
galaxies observed by Vega Beltr\'an et al. (2000).  All these galaxies 
are bright ($B_T\leq13.5$) and nearby objects ($V_\odot < 5800$ \kms) 
with an intermediate-to-high inclination ($i \geq 45 \dg$) and their 
Hubble morphological types run from S0 to Sc. 
 
Since the presence of ionized-gas supported by non-circular motions 
have been observed only in lenticular or bulge-dominated spiral 
galaxies, we focused our attention on S0 -- Sb galaxies.  We selected 
unbarred galaxies according to the classification of both Sandage \& 
Tammann (1981, hereafter RSA) and de Vaucouleurs et al. (1991, 
hereafter RC3). To better disentangle the contributions of the bulge, 
the disc and eventually of the dark matter halo to the total mass of 
the galaxy we choosed between the sample objects those with the more 
extended ionized-gas and stellar kinematics (in units of $R_{25}$). At 
the end of this selection process we remained with five galaxies: an 
S0 (NGC~980), two Sa (NGC~772 and NGC~5064) and two Sb spirals 
(NGC~3898 and NGC~7782). 
 
All the selected galaxies display a smooth and regular morphology, 
except for NGC 772 which is characterized by a strong lopsidedness 
(Fig.~\ref{fig:n772_Halpha}). However the symmetry of velocity curves 
and velocity-dispersion profiles of both ionized gas and stars 
(Fig.~\ref{fig:n772_kinematics}) suggests us that also NGC 772 has an 
axisymmetric structure at least in the radial region were the 
kinematic parameters were measured (corresponding to $0.2\,R_{25}$). 
 
An accurate analysis of the photometric and kinematic properties of 
the selected galaxies (see Vega Beltr\`an 1999 for details)  
showed that the surface 
brightness of NGC~980 is characterized by a strong twisting of the 
isophotes and that the LOSVD profiles of NGC 5064 have a 
strongly non-Gaussian shape. We interpreted these two phenomena as 
due to the misalignment between bulge and disc of NGC~980 (or even to the 
presence of a triaxial bulge) and to the possible coexistence 
in NGC~5064 of two counterrotating components, respectively. 
Therefore the two galaxies were discarded and are not studied here. 
 
An overview of the properties of the remaining galaxies, namely NGC 772, 
NGC 3898 and NGC 7782 which will be the subject of our investigation, 
is given in Table \ref{tab:properties}. 
Their available photometric and kinematical data are listed here briefly.

\begin{table*}    
\caption{Optical and radio properties of NGC 772, NGC 3898 and NGC 
7782}    
\begin{center}    
\begin{tabular}{llll}    
\hline    
\noalign{\smallskip}     
Parameter & NGC 772 & NGC 3898 & NGC 7782 \\    
\noalign{\smallskip}     
\hline    
\noalign{\smallskip}     
Other name  &  
 UGC 1466; PGC 7525 & UGC 6787; PGC 36921 & UGC 12834; PGC 72788 \\    
Morphological type  & 
 Sb(rs)I$\;^{\rm a}$; Sb$\;^{\rm b}$; SAS3$\;^{\rm c}$ &    
 SaI$\;^{\rm a}$; Sa$\;^{\rm b}$; SAS2$\;^{\rm c}$ &    
 Sb(s)I-II$\;^{\rm a}$; Sb$\;^{\rm b}$; SAS3$\;^{\rm c}$ \\     
Heliocentric systemic velocity (\kms)$\;^{\rm d}$ & 
 $2470\pm10$ & $1184\pm10$ & $5430\pm10$ \\    
Distance (Mpc)$\;^{\rm d}$ & 34.7 &  17.1 & 75.2\\    
Major-axis position angle$\;^{\rm c}$ & 
 $130^\circ$ & $107^\circ$ & $175^\circ$\\    
Apparent isophotal diameters $\;^{\rm c}$   & 
 $7\farcm2\times4\farcm3$  & $4\farcm4\times2\farcm6$   &  
 $2\farcs4\times1\farcs3$ \\     
Inclination$\;^{\rm d}$ & 
 $54^\circ$ & $54^\circ$ & $58^\circ$\\    
Apparent $V_T$ magnitude (mag)$\;^{\rm c}$  & 
 10.31 & 10.70 & 12.23\\    
Total $(B-V)_T$ color index (mag)$\;^{\rm c}$           &  0.78      &  
0.90      &  0.85\\     
Total corrected $V$ luminosity $L_{V^0_T}$ (\lvsun)$\;^{\rm d}$              
  & $11.2\cdot10^{10}$ & $1.6\cdot10^{10}$ & $9.7\cdot10^{10}$\\    
\hi\ linewidth at $20\%$ of the peak (\kms)  & 473$\;^{\rm e}$ & 
504$\;^{\rm f}$ & 570$\;^{\rm g}$ \\    
\hi\ linewidth at $50\%$ of the peak (\kms)  & 410$\;^{\rm e}$ & 
469$\;^{\rm f}$ & 559$\;^{\rm g}$ \\    
Mass of neutral hydrogen $M_{\rm HI}$ (\msun) & 
 $25.6\cdot10^{9}$$\;^{\rm e}$ & $2.6\cdot10^9$$\;^{\rm f}$ & 
 $14.5\cdot10^9$$\;^{\rm g}$\\  
Mass of cool dust $M_{\rm d}$ (\msun)$\;^{\rm h}$ & 
 $2.3\cdot10^7$ & $0.07\cdot10^7$ & $2.3\cdot10^7$\\    
\noalign{\smallskip}     
\hline    
\noalign{\smallskip}     
\end{tabular}    
\begin{minipage}{16cm}    
$^{\rm a}$ from RSA.\\    
$^{\rm b}$ from Nilson (1973, hereafter UGC).\\    
$^{\rm c}$ from RC3.      
The apparent isophotal diameters are measured at    
a surface brightness level of $\mu_B = 25$ $\rm mag\;arcsec^{-2}$.\\     
$^{\rm d}$ from this paper. The distance is derived as $V_0/H_0$ with    
$V_0$ the velocity relative to the centroid of the Local Group    
obtained from the heliocentric systemic velocity as in RSA 
and $H_0 = 75$ \kms\ Mpc$^{-1}$. The inclination $i$ is    
derived as $\cos ^{2} i = (q^2-q_0^2)/(1-q_0^2)$, where the observed    
axial ratio is taken from RC3 and    
an intrinsic flattening of $q_0 = 0.11$ has been assumed following    
Guthrie (1992).\\  
$^{\rm e}$ from Rhee \& van Albada (1996), the neutral hydrogen  
     mass has been scaled for the adopted distance.\\    
$^{\rm f}$ from van Driel \& van Woerden (1994), the neutral hydrogen  
     mass has been scaled for the adopted distance.\\ 
$^{\rm g}$ from Krumm \& Salpeter (1980), the neutral hydrogen  
     mass has been scaled for the adopted distance.\\ 
$^{\rm h}$ derived following Young et al. (1989) from the     
IRAS flux densities at 60 and 100 $\mu$m (Moshir et al. 1990).\\    
    
\end{minipage}    
\end{center}    
\label{tab:properties}    
\end{table*}    
    
\subsection{NGC 772}    
\label{sec:n772_properties}    
    
Surface photometry of NGC 772 was obtained in the $B$ band by Lu 
(1998), who also derived the photometric parameters of the exponential 
disc, and in the $V$ and $I$ bands by Heraudeau \& Simien (1996).  NGC 
772 belongs to the sample of 34 early-to-late spiral galaxies, whose 
major-axis stellar velocities has been recently measured by Heraudeau 
\& Simien (1998). 
NGC 772 was observed in the 21-cm line of neutral 
hydrogen by Rhee \& van Albada (1996), who obtained the 
\hi\ position-velocity map, global velocity profile and 
radial surface density distribution. These data show the asymmetric 
distribution of the \hi\ at radii larger than $5'$ from the centre. On 
the NW side a low rotation-velocity ($V_{HI} \simeq 100$ \kms) 
component can be traced out to $9'$, while on the SE the \hi\ emission 
extends to $5'$ with a rotation velocity of $260$ \kms . 
    
\subsection{NGC 3898}    
\label{sec:n3898_properties}    
        
Surface photometry of NGC 3898 is available in $B$ (Barbon, Benacchio 
\& Capaccioli 1978; Boroson 1981), $V$ (Watanabe 1983; Kodaira, 
Okamura \& Ichikawa 1990; Heraudeau \& Simien 1996), $I$ (Heraudeau \& 
Simien 1996), $r$ (Kent 1988), $J$ (Giovanardi \& Hunt 1996; Moriondo, 
Giovanardi \& Hunt 1998a), $H$ (Giovanardi \& Hunt 1996), and $K$ band 
(Giovanardi \& Hunt 1996; Moriondo et al. 1998a). 
Whitmore, Rubin \& Ford (1984), Fillmore et al.  (1986), Heraudeau et 
al. (1999) measured the major-axis stellar velocity curve and 
velocity-dispersion profile of NGC 3898. 
The ionized-gas rotation curve was 
obtained along the galaxy major axis by  
Rubin et al. (1985) and Fillmore et al. (1986).
The distribution and velocity field of \hi\ were studied in detail by 
van Driel \& van Woerden (1994), who also derived using the 
maximum-disc hypothesis (van Albada \& Sancisi 1986) the mass 
contribution of bulge, disc and dark halo by fitting the combined 
\ha/\hi\ rotation curve using the photometric parameters of bulge and 
disc obtained from Watanabe's (1983) surface-brightness profile. 
Other mass models for NGC 3898 have been obtained by Fillmore et 
al. (1986) using both gas and stellar kinematics and by Kent (1988) 
and Moriondo, Giovanardi \& Hunt (1998b)  
who adopted the ionized-gas kinematics by Rubin et al. (1985).

\subsection{NGC 7782}    
\label{sec:n7782_properties}    
        
Surface photometry of NGC 7782 has been obtained in the $V$ band 
(Kodaira et al. 1990), the $r$ band (Courteau 1996) and the $H$ band 
(Moriondo et al. 1999). The only available bulge-disc decomposition 
for NGC 7782 is that of Baggett, Baggett \& Anderson (1998) based on 
the data of Kodaira et al. (1990). No spatially resolved kinematics for 
the gaseous and stellar components have been obtained for this galaxy 
either at optical or at radio wavelengths. 
    
\section{Observations and data reduction}    
\label{sec:observations}    
    
\subsection{Long-slit spectroscopy}    
\label{sec:spectroscopy}    
    
The spectroscopic observations of NGC 772, NGC 3898 and 
NGC 7782 were carried out in two different runs during 
October and December 1990 at the 4.5-m Multiple Mirror 
Telescope (MMT) on Mt. Hopkins (Arizona, 
USA). The 1200 $\rm grooves\;mm^{-1}$ grating blazed at 
5767 \AA \ was used in the first order in combination 
with a $1\farcs25\times3\farcm0$ slit and the Loral 
$1200\;\times\;800$ CCD with pixels of $15\;\times\;15$ 
$\rm \mu m^2$.  It yielded a wavelength coverage of 650 
\AA\ between 4850 and 5500 \AA\ with a reciprocal 
dispersion of 54.7 $\rm \AA\;mm^{-1}$.  No on-chip 
binning was performed and every spectrum pixel 
corresponded to 0.82 \AA\ by $0\farcs30$. 
    
NGC 3898 was also observed at the Isaac Newton Telescope 
(INT) in La Palma (Spain) on March 19, 1996.  The 
Intermediate Dispersion Spectrograph (IDS) was used with 
a $1\farcs9 \times 4\farcm0$ slit, the 500~mm camera, 
the AgRed collimator, and the H1800V grating with 1800 
grooves$\rm\;mm^{-1}$ at first order.  This 
instrumental set-up yielded a wavelength coverage of 
$240$ \AA\ between 6650 \AA\ and 6890 \AA\ with a 
reciprocal dispersion of 9.92~$\rm\AA\;mm^{-1}$. No 
on-chip binning was applied on the adopted 1024$\times$1024 
TK1024A CCD. Each 24 $\mu$m~$\times$ 24~$\mu$m spectrum 
pixel corresponds to 0.24 \AA\ by $0\farcs33$.

\begin{table}    
\caption{Log of the spectroscopic observations}    
\begin{center}    
\begin{tabular}{l c c c r}    
\hline    
\noalign{\smallskip}    
\multicolumn{1}{c}{Object} &    
\multicolumn{1}{c}{Date} &    
\multicolumn{1}{c}{Telescope} &    
\multicolumn{1}{c}{t$_{\it exp}$} &    
\multicolumn{1}{c}{P.A.} \\    
\noalign{\smallskip}    
\multicolumn{1}{c}{} &    
\multicolumn{1}{c}{} &    
\multicolumn{1}{c}{} &    
\multicolumn{1}{c}{[s]} &    
\multicolumn{1}{c}{[$^\circ$]} \\    
\noalign{\smallskip}    
\hline    
\noalign{\smallskip}    
NGC~772  & 22 Oct 1990 & MMT & $3600$        & 130\\    
NGC~3898 & 18 Dec 1990 & MMT & $3600$        & 107\\    
         & 19 Mar 1996 & INT & $4\times3600$ & 107\\    
NGC~7782 & 22 Oct 1990 & MMT & $3600$        & 175\\    
\noalign{\smallskip}    
\hline    
\label{tab:log_spectroscopy}    
\end{tabular}    
\end{center}    
\end{table}    
    
At the beginning of each exposure, the slit was centred 
on the galaxy nucleus using the guiding TV camera and 
aligned along the galaxy major axis. The details on the 
slit position and spectra exposure times are given in 
Table~\ref{tab:log_spectroscopy}.  At the MMT spectra of 
a number of late-G and early-K giant stars were obtained 
with the same set up to serve as templates in measuring 
the stellar kinematics. In all observing runs, comparison 
exposures of the arc lamp were taken before and after 
each object integration to allow an accurate wavelength 
calibration. Quartz-lamp and twilight-sky flat fields 
were used to map pixel-to-pixel sensitivity variations 
and large-scale illumination patterns.  The seeing during 
the observations was typically between $1\farcs2$ and 
$1\farcs5$ FWHM. 
    
Using standard MIDAS\footnote{MIDAS is developed and 
maintained by the European Southern Observatory} routines 
the spectra were bias subtracted, flat-field corrected 
and wavelength calibrated.  Cosmic rays were identified 
by comparing the counts in each pixel with the local mean 
and standard deviation (as obtained by the Poisson 
statistics of the photons knowing the gain and readout 
noise of the detector), and corrected by 
interpolating. The instrumental resolution was derived by 
measuring the Gaussian FWHM of a dozen of unblended 
arc-lamp lines distributed over the whole spectral range 
of a wavelength-calibrated comparison spectrum.  We found 
a mean value of $\rm FWHM = 2.24 \pm 0.26 $ \AA\ and of $\rm FWHM 
= 2.57 \pm 0.11$ \AA\ for the MMT spectra obtained in October 
1990 and December 1990, respectively. They correspond to 
an instrumental velocity dispersion of $\sigma_{\it instr} = 55$  
\kms\ and  $\sigma_{\it instr} = 64$ \kms\ at 5150 \AA.   
For the INT spectra we measured a 
mean $\rm FWHM = 0.87 \pm 0.04$ \AA\ that, in the range of 
the observed gas emission lines, corresponds to an 
instrumental velocity dispersion of 
 $\sigma_{\it instr} = 17$~\kms . 
    
The stellar kinematic parameters were measured from the
absorption lines present on MMT spectra using the Fourier
Correlation Quotient method (Bender 1990), as applied by
Bender, Saglia \& Gerhard (1994). The spectra of the
G5III star HR 7778 and K2III star HR~6415 provided the
best match to galaxy spectra obtained in October and
December 1990, respectively, so they were used as
templates to measure the galaxy stellar velocities in the
two runs.  The stellar kinematics of NGC 772, NGC 3898
and NGC 7782 are discussed in Sects.
\ref{sec:n772_kinematics}, \ref{sec:n3898_kinematics},
and \ref{sec:n7782_kinematics}, and the key parameters
are tabulated in Tabs. 4, 5 and 6.  Each table provides
the radial distance from the galaxy centre in arcsec, the
observed heliocentric velocity and the velocity
dispersion in \kms, and the Gauss-Hermite coefficients
$h_3$ and $h_4$.
    
The ionized-gas kinematics was derived by measuring the position and 
width of the \oiii\ and \ha\ emission lines in the MMT and INT 
spectra, respectively. Using the MIDAS package ALICE we fitted 
interactively a Gaussian to the emission line and a polynomial to its 
surrounding continuum. The central wavelength and FWHM (corrected for 
instrumental FWHM) of the fitting Gaussian were converted into radial 
velocity and velocity dispersion, respectively. The resulting 
velocities were corrected to the heliocentric frame of reference. At 
radii where the intensity of the relevant emission was low, we 
averaged from 3 to 7 adjacent spectral rows to improve the 
signal-to-noise ratio of the line.  The ionized-gas kinematics of NGC 
772, NGC 3898 and NGC 7782 are described in 
Sects. \ref{sec:n772_kinematics}, \ref{sec:n3898_kinematics}, and 
\ref{sec:n7782_kinematics}, and the key parameters are 
tabulated in Tabs. 7, 8 and 9. Each table provides the radial 
distance from the galaxy centre in arcsec, the \oiii\ (and 
the \ha\ only for NGC 3898) observed heliocentric velocity and 
velocity dispersion in \kms. 
    
\medskip    
\noindent    
    
\subsection{Broad-band imaging}    
\label{sec:broadimaging}    
    
The broad-band imaging of the three galaxies was carried 
out at the 1.83-m Vatican Advanced Technology Telescope 
(VATT) operated in the Mt. Graham International 
Observatory (Arizona, USA) in two observing runs on March 
and November 1997. A back-illuminated $2048\times2048$ 
Loral CCD with $15\times15$ $\mu$m$^{2}$ pixels was used 
as detector at the aplanatic Gregorian focus (f/9) of the 
telescope.  It yielded a field of view of 
$6\farcm4\times6\farcm4$ with an image scale of 
$0\farcs4$ pixel$^{-1}$ after a $2\times2$ pixel binning. 
The gain and the readout noise were 1.4 e$^-$ ADU$^{-1}$ 
and 6.5 e$^-$, respectively. 
    
At regular intervals during each night, different bias 
frames (typically ten) were taken to check possible 
slight bias-level variations.  All galaxies were observed 
twice in the Johnson $V$ band.  The date and the duration 
of all the exposures are given in 
Table~\ref{tab:log_imaging}.  A number of twilight sky 
flats were taken at the beginning and at the end of 
the nights.  No photometric standard was observed. 
    
\begin{table}    
\caption{Log of the broad and narrow-band observations}    
\begin{center}    
\begin{tabular}{l r c r c}    
\hline    
\noalign{\smallskip}    
\multicolumn{1}{c}{Object} &    
\multicolumn{1}{c}{Date} &    
\multicolumn{1}{c}{Filter} &    
\multicolumn{1}{c}{t$_{\it exp}$} &    
\multicolumn{1}{c}{Seeing$^{\rm a}$} \\    
\noalign{\smallskip}    
\multicolumn{1}{c}{} &    
\multicolumn{1}{c}{} &    
\multicolumn{1}{c}{} &    
\multicolumn{1}{c}{[s]} &    
\multicolumn{1}{c}{[$''$]} \\    
\noalign{\smallskip}    
\hline    
\noalign{\smallskip}    
NGC~772  & 02 Nov 1997 & $V$   & $2\times180$ & 1.7\\    
         & 07 Nov 1997 & r6450 & $4\times600$ & 1.9\\    
         & 07 Nov 1997 & r6630 & $4\times600$ & 1.8\\    
NGC~3898 & 11 Mar 1997 & $V$   & $2\times60 $ & 1.4\\    
         & 11 Mar 1997 & r6450 & $4\times540,\,2\times600$ & 1.3\\    
         & 11 Mar 1997 & r6580 & $2\times500,\,2\times540$ & 1.3\\    
NGC~7782 & 02 Nov 1997 & $V$   & $2\times360$ & 2.2\\    
         & 10 Nov 1997 & r6450 & $3\times600$ & 1.8\\    
         & 10 Nov 1997 & r6680 & $3\times600$ & 2.2\\    
\noalign{\smallskip}    
\hline    
\noalign{\smallskip}    
\label{tab:log_imaging}    
\end{tabular}    
\begin{minipage}{8.cm}    
$^{\rm a}$ Seeing FWHM measured on the final resulting frame.\\    
\end{minipage}    
\end{center}    
\end{table}    
    
The data reduction was carried out using standard 
IRAF\footnote{IRAF is distributed by the National Optical 
Astronomy Observatories which are operated by the 
Association of Universities for Research in Astronomy 
(AURA) under cooperative agreement with the National 
Science Foundation.} routines. 
All the frames were bias subtracted and corrected for 
pixel-to-pixel intensity variations by using a mean flat 
field for each night.  The different frames of each 
galaxy were shifted and aligned to an accuracy of a few 
hundredths of a pixel using common field stars as 
reference.  After checking that their point spread 
functions (PSF) were comparable, the frames were averaged 
to obtain a single $V-$band image. The cosmic rays were 
identified and removed during the averaging routine. 
Two-dimensional Gaussian fits to the field stars in the 
resulting images yielded the final FWHM measurement of 
seeing PSF listed in Table~\ref{tab:log_imaging}. 
In each final frame the mean value of the sky level was 
determined in a large number of $5\times5$ pixel 
areas. These areas were selected in empty regions of the 
frames, which were free of objects and far from the 
galaxy to avoid the contamination of the light of field 
stars and galaxies as well as of the target galaxy 
itself.  The sky value of each frame is the average of 
these mean values.  For the estimate of error in the sky 
determination we adopted half the difference between 
the maximum and minimum of the mean values obtained for 
the small areas.  
    
\begin{figure}    
\vspace*{9cm}    
\includegraphics{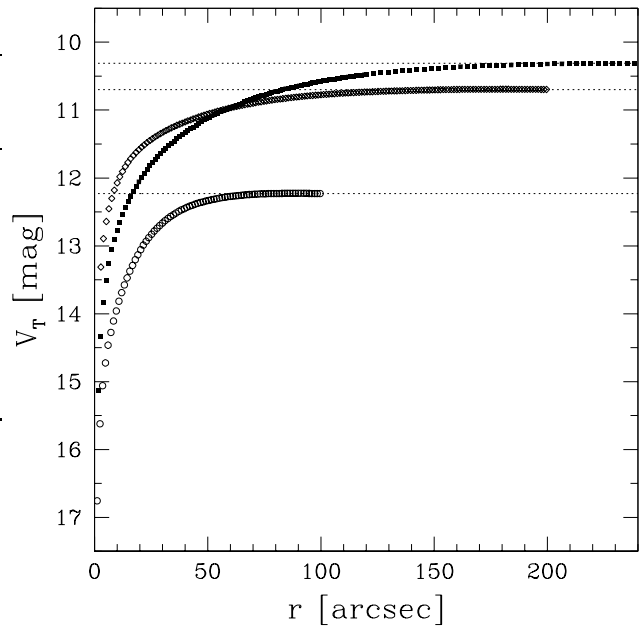}    
\caption[]{Calibrated luminosity growth curves of NGC~772     
  ({\it filled squares\/}, $V_T =10.31$ mag), NGC~3898 ({\it open    
  diamonds\/}, $V_T =10.70$ mag) and NGC~7782 ({\it open circles\/},    
  $V_T =12.23$ mag).  For each galaxy the {\it dashed line\/} shows    
  the asymptotic magnitude taken from RC3.}    
\label{fig:growthcurves}    
\end{figure}

For each galaxy we derived a `luminosity growth curve' 
by measuring the integrated magnitudes within circular 
apertures of increasing radius by means of the IRAF task 
ELLIPSE within the STSDAS package. 
Absolute calibration was performed by fitting 
the constant portion of the growth curves 
(Fig.~\ref{fig:growthcurves}) to the corresponding total 
magnitude $V_T$ given by RC3.

Before fitting ellipses to the galaxy images, we removed 
the field stars using the IRAF routines within the 
DAOPHOT package.  Special care was taken to remove 
satured stars and field galaxies which were edited out by 
hand, and removed by replacing them with the local 
average of counts. When bright stars were too close to 
the galaxy their light was masked out to where the galaxy 
light started to be dominant. 
For each galaxy we derived the surface-brightness profile 
and isophotal shape parameters by ellipse fitting to its 
isophotes using the isophote fitting program ELLIPSE 
(see Jedrzejewky 1987 for 
details on the fitting procedure). The resulting surface 
brightness, ellipticity, position angle and 
$\cos{4\theta}$ Fourier coefficient radial profiles 
for NGC 772, NGC 3898, and NGC 7782 are 
presented in Sects. \ref{sec:n772_photometry}, 
\ref{sec:n3898_photometry}, and 
\ref{sec:n7782_photometry}, and tabulated in 
Tabs. 10, 11, and 12.
Each table provides the isophotal semi-major 
axis in arcsec, surface brightness in 
\mas, ellipticity, position angle in 
degrees, and the $\cos{4\theta}$ Fourier coefficient.

\subsection{Narrow-band imaging}    
\label{sec:narrowimaging}

The narrow-band \ha\ imaging of NGC~772, NGC~3898 and 
NGC~7782 was performed at the VATT in the same observing 
runs, during which we obtained the $V-$band images. 
    
For each galaxy we obtained at least two emission-band and two    
continuum-band images using the set of interference filters kindly    
provided by R.C. Kennicutt.    
The emission-band images were taken with three different 
interference filters ($\lambda_c = 6630$ \AA, $\lambda_c 
= 6580$ \AA, and $\lambda_c = 6680$ \AA\ respectively for 
NGC~772, NGC~3898 and NGC~7782; $\Delta\lambda_{\rm FWHM} 
= 70$ \AA) in order to isolate the spectral region 
characterized by the redshifted \ha\ and \nii\ 
($\lambda\lambda$6548.0, 6583.4 \AA) emission lines 
according to galaxy systemic velocity.  The 
continuum-band images were taken through an interference 
filter ($\lambda_c = 6450$ \AA; $\Delta\lambda_{\rm FWHM} 
= 70$ \AA), which has been selected to observe an 
emission-free spectral region sufficiently near to that 
of the emission-band filters in order to subtract off the 
stellar continuum in the emission-band images. Duration, 
date and filter of all the in-band and off-band exposures 
are collected in Table~\ref{tab:log_imaging}. Flat-field 
exposures of the twilight sky were taken at dawn and 
sunset for each filter. 
    
The narrow-band images of the target galaxies were bias 
subtracted, flat-field corrected, aligned, averaged and 
cleaned of cosmic ray events in the same way as for the 
broad-band images.  As result of this standard data 
reduction we obtained for each galaxy a single 
emission-band image and a single continuum-band 
image. The seeing FWHM's of these images as measured by 
fitting a two-dimensional Gaussian to the field stars are 
listed in Table~\ref{tab:log_imaging}.  For each galaxy 
the best-seeing image was convolved with a Gaussian PSF 
to yield the same PSF FWHM of the worst-seeing image. 
Finally the continuum-band image was suitably scaled and 
subtracted from the emission-band image to obtain a 
continuum-free map of the \ha$+$\nii\ emission of the 
galaxy. The scale factor for the continuum-band image was 
estimated by comparing the intensity a large number of 
field stars in the two bandpasses.  The resulting 
continuum-band and continuum-free images of NGC~772, 
NGC~3898 and NGC~7782 are given in 
Sects. \ref{sec:n772_Halpha}, \ref{sec:n3898_Halpha}, and 
\ref{sec:n7782_Halpha}.

\section{Structure models}    
\label{sec:structure}    
    
In order to describe the mass structure of each galaxy, we have to
take advantage of both the photometric and kinematic data available.
    
We describe each galaxy by the superposition of two 
different components, namely a bulge and a disc.  Both 
the components are assumed to be oblate, with the 
isodensity surfaces being similar concentric 
spheroids. In this framework, what we call a `disc' is 
in fact a spheroidal component with a very high 
ellipticity, which we assume to be constant with radius. 
    
Once the luminosity density parameters are constrained by 
the photometry (Sect.~\ref{sec:decomposition}), we 
use the kinematics to evaluate the mass-to-light ratios and to 
discuss the gas velocity distribution. 
    
\subsection{Photometric decomposition}    
\label{sec:decomposition}    
    
Conventional bulge-disc photometric decompositions based 
on elliptically averaged surface-brightness profiles are 
subject to strong systematic errors due to the different 
intrinsic shapes of bulge and disc and to the viewing 
angle of the galaxy (see Byun \& Freeman 1995 for an 
extensive discussion). 
For this reason we applied to the $V-$band images of NGC 
772, NGC 3898 and NGC 7782 a two--dimensional parametric 
decomposition technique. The bulge and disc photometric 
parameters used as the initial trial in modelling the 
observed kinematics were derived by fitting directly 
the galaxy surface-brightness distributions.  Our 
decomposition method is based on the technique developed 
by Byun \& Freeman (1995) which we improved by 
introducing convolution with the seeing PSF and by 
weighting the observed surface brightness measured in 
each pixel according to the associated Poissonian noise and CCD 
readout noise. 
    
\subsubsection{Two-dimensional bulge-disc parametric decomposition}    
    
We assumed the galaxy surface-brightness distribution to be the    
sum of the contributions of an oblate bulge and an infinitely thin disc.    
We adopted the $r^{1/4}$ law (de Vaucouleurs 1948) to describe the    
surface brightness of the bulge component    
    
\begin{eqnarray}    
\lefteqn{I_{\it bulge}\,(x,y) = } \nonumber \\    
  & & I_e\,\exp{\left\{-7.67\left[    
  \left(\frac{\sqrt{x^2+y^2(b/a)_{\it bulge}^{-2}}}{r_e}\right)^{1/4}    
  -1\right]\right\}}    
\label{eqn:bulge_sufbright}    
\end{eqnarray}    
    
\noindent     
where $I_e$ and $r_e$ are respectively the effective surface    
brightness and effective radius of the bulge.      
$x$ and $y$ are the apparent distances from galactic centre along the    
major and minor axis, and $(b/a)_{\it bulge}$ is the apparent axial ratio    
of bulge. Bulge isophotes are ellipses with constant apparent    
ellipticity $\epsilon_{\it bulge} = 1 - (b/a)_{\it bulge}$.    
    
The disc component is assumed to have an exponential surface    
brightness (Freeman 1970)     
    
\begin{equation}    
  I_{\it disc}\,(x,y) =     
  I_0\,\exp{\left[\frac{\sqrt{x^2+y^2(b/a)_{\it disc}^{-2}}}{h} \right]}    
\label{eqn:disc_surfbright}    
\end{equation}    
    
\noindent     
where $\mu_0$ and $h$ are respectively the disc central surface    
brightness and scale length. $x$ and $y$ the apparent distances from    
galactic centre along the major and minor axis, and $(b/a)_{\it disc}$    
is the apparent axial ratio of disc. Disc isophotes are ellipses with    
constant apparent ellipticity $\epsilon_{\it disc} = 1 - (b/a)_{\it    
disc}$ and the disc inclination is $i = \arccos{(b/a)_{disc}}$.    
    
\setcounter{table}{12}    
\begin{table*}       
\caption{Parameters from the bulge-disc photometric decomposition}    
\begin{center}    
\begin{tabular}{lcrccccccccc}    
\noalign{\smallskip}    
\hline    
\noalign{\smallskip}    
\multicolumn{1}{c}{Object} &     
\multicolumn{4}{c}{Bulge} &    
\multicolumn{1}{c}{} &     
\multicolumn{4}{c}{Disc} &    
\multicolumn{1}{c}{}  &    
\multicolumn{1}{c}{Fit$^{\rm a}$}  \\    
\noalign{\smallskip}    
\cline{2-5}\cline{7-10}    
\noalign{\smallskip}    
\multicolumn{1}{c}{} &    
\multicolumn{1}{c}{$\mu_{e}$} &    
\multicolumn{2}{c}{$r_{e}$} &    
\multicolumn{1}{c}{$(b/a)_{\it bulge}$} &    
\multicolumn{1}{c}{}&    
\multicolumn{1}{c}{$\mu_{0}$} &    
\multicolumn{2}{c}{$h$} &    
\multicolumn{1}{c}{$(b/a)_{\it disc}$} &    
\multicolumn{1}{c}{i} &    
\multicolumn{1}{c}{}\\    
\multicolumn{1}{c}{} &    
\multicolumn{1}{c}{[\mas]} &    
\multicolumn{1}{c}{[$''$]} &    
\multicolumn{1}{c}{[kpc]} &    
\multicolumn{1}{c}{} &    
\multicolumn{1}{c}{} &    
\multicolumn{1}{c}{[\mas]} &    
\multicolumn{1}{c}{[$''$]} &    
\multicolumn{1}{c}{[kpc]} &    
\multicolumn{1}{c}{} &    
\multicolumn{1}{c}{[$^{\circ}$]} &    
\multicolumn{1}{c}{}\\    
\noalign{\smallskip}    
\hline    
\noalign{\smallskip}    
NGC~772 & 22.7  & 60.5 & 10.2 & 0.66 && 21.5 & 96.3 & 16.2 & 0.39 & 67 & 1D\\  
        & 22.8  & 65.4 & 11.0 & 0.77 && 21.3 & 67.0 & 11.3 & 0.62 & 52 & 2D\\  
NGC~3898& 20.9  & 25.6 &  2.1 & 0.64 && 21.6 & 45.9 &  3.8 & 0.37 & 68 & 1D\\  
        & 20.6  & 18.9 &  1.6 & 0.71 && 20.4 & 29.0 &  2.4 & 0.52 & 59 & 2D\\  
NGC~7782$^{\rm b}$     
        & 20.2  &  3.6 &  1.3 & 0.85 && 20.5 & 23.3 &  8.5 & 0.41 & 66 & 1D\\ 
        & 19.9  &  2.9 &  1.1 & 0.71 && 20.3 & 21.6 &  7.8 & 0.56 & 56 & 2D\\  
\hline    
\label{tab:result2Dfit}    
\end{tabular}    
\begin{minipage}{16.cm} 
$^{\rm a}$ 1D = parametric fit on the ellipse-averaged profiles, 
           2D = two-dimensional parametric fit.\\   
$^{\rm b}$ For NGC 7782 the best-fit to the observed 
surface-brightness distribution has been obtained adopting an 
exponential bulge (see discussion in Sec.~5.3.4).   
\end{minipage}    
\end{center}    
\end{table*}    
    
To derive the photometric parameters of the bulge ($I_e$, $r_e$ and 
$(b/a)_{bulge}$) and the disc ($I_0$, $h$ and $(b/a)_{disc}$) we 
fitted iteratively a model of the surface brightness to the 
observations using a non-linear $\chi^2$ minimization based on the 
Levenberg-Marquardt method (e.g. Bevington \& Robinson 1992; Press et 
al. 1996). 
    
The galaxy centre, the bulge and disc position angles, as well as the    
sky level are not free parameters in our decomposition.  The galaxy    
centre is assumed to be the mean centre of the ellipses which fit the galaxy    
isophotes. Bulge and disc are assumed to have the same position angle    
which is the mean value of the major-axis position angle of ellipses    
which fit the galaxy isophotes.  The sky level is estimated as described in    
Sect. \ref{sec:observations}.    
The seeing effects were taken into account by convolving the model 
image with a circular Gaussian PSF with the FWHM measured using the 
stars in the image (Table~\ref{tab:log_imaging}). The convolution was 
performed as a product in Fourier domain before the $\chi^2$ minimization. 
    
For each pixel $(x,y)$, the observed galaxy photon counts $\cal I_{\it 
gal}$ are compared with those predicted from the model $\cal I_{\it 
bulge} + I_{\it disc}$.  Each pixel is weighted according to the 
variance of its total observed photon counts due to 
the contribution of both galaxy and sky, and determined assuming 
photon noise limitation.  Therefore we can 
write 
    
\begin{equation}    
  \chi^{2} = \sum_{x=1}^{N} \sum_{y=1}^{M}       
  \frac{[\,{\cal I}_{\it mod}\,(x,y) -    
  {\cal I}_{\it gal}\,(x,y)\,]^2}{{\cal I}_{\it gal}\,(x,y) + 
  {\cal I}_{\it sky}\,(x,y)}    
\label{eqn:chi2}    
\end{equation}    
    
\noindent    
with $x$ and $y$ ranging over the whole $N \times M$ pixel image.    
  
To derive the six free parameters of the model surface-brightness 
distribution, we adopted as the initial trial for $\chi^2$ 
minimization the values obtained by performing a standard photometric 
decomposition with a parametric technique similar to that adopted by 
Kent (1985). 
In fact we decomposed the observed surface-brightness profile along
both the major axis (obtained by fitting ellipses to isophotes) and
the minor axis (obtained from the major-axis profile scaled by the
factor $1-\epsilon$) as the sum of an $r^{1/4}$ oblate bulge plus an
exponential infinitely thin disc.  To take into account the seeing
effect, we truncated the major and minor-axis profiles at a radius
corresponding to a couple of PSF FWHM from the centre.  We assumed the
minor-axis profiles of bulge and disc to be the same as the major-axis
profile scaled respectively by a factor $1-\epsilon_{\it bulge} =
(b/a)_{\it bulge}$ and $1-\epsilon_{\it disc} = (b/a)_{\it disc}$.  A
least-squares fit of the model to the photometric data provided the
values of $I_e$, $r_e$, $(b/a)_{\it bulge}$, $I_0$, $h$ and
$(b/a)_{\it disc}$ to be used as the initial trial parameters for the
two-dimensional photometric decomposition.
    
The photometric parameters of the bulge and the disc obtained for NGC
772, NGC 3898 and NGC 7782 are given in Table~\ref{tab:result2Dfit}.
The model surface brightnesses obtained for the sample galaxies are
discussed in Sects. \ref{sec:n772_photometry},
\ref{sec:n3898_photometry}, and \ref{sec:n7782_photometry}. 
    
\subsubsection{Test with model galaxies}    
    
To test the reliability and accuracy of our two-dimensional technique    
for bulge-disc photometric decomposition we applied the decomposition    
program to a set of artificial disc galaxies.  We generated 100 images    
of galaxies with an $r^{1/4}$ oblate bulge plus an exponential    
infinitely thin disc.  The scale surface-brightness, scale length, and    
apparent axial ratios of bulge and disc of the artificial galaxies    
were randomly chosen in the range of values observed by Kent (1985)    
for a large sample of S0 and spiral galaxies. The adopted ranges are    
    
\begin{equation}    
18 \leq \mu_e \leq 24\  {\rm mag\;arcsec}^{-2},\\    
\end{equation}    
    
\begin{equation}    
3 \leq r_e \leq 10\ {\rm kpc}, \\    
\end{equation}    
    
\begin{equation}    
0.6 \leq (b/a)_{\it bulge} \leq 1,    
\end{equation}    
    
\noindent    
for the bulge parameters, and    
    
\begin{equation}    
19 \leq \mu_e \leq 23\  {\rm mag\;arcsec}^{-2},\\    
\end{equation}    
    
\begin{equation}    
3 \leq r_e \leq 15\ {\rm kpc}, \\    
\end{equation}    
    
\begin{equation}    
0.2 \leq (b/a)_{\it disc} \leq 1    
\end{equation}    
    
\noindent    
for the disc parameters. The parameters of the artificial galaxies    
have also to satisfy the following conditions    
    
\begin{equation}    
(b/a)_{\it disc} < (b/a)_{\it bulge},     
\end{equation}    
    
\begin{equation}    
0 < B/T < 0.8.    
\end{equation}    
    
All the simulated galaxies were assumed to be at the Virgo Cluster    
distance ($d=17$ Mpc, Freedman et al. 1994) corresponding to a scale of    
$82.4$ pc arcsec$^{-1}$. The pixel scale used was $0\farcs4$    
pixel$^{-1}$ and the CCD gain and readout noise were respectively 1.4    
e$^-$ ADU$^{-1}$ and 6.5 e$^-$ in order to simulate the VATT    
observations. We fixed the seeing FWHM at $2\farcs0$.  We added a sky    
background level of $10$ counts pixel$^{-1}$ which corresponds to a    
sky surface brightness of 21.8 \mas\ in the $V-$band adopting the    
absolute calibration derived from our real observations.  This is a    
typical surface brightness value for a dark sky as reported by Binney    
\& Merrifield (1998). An appropriate level of noise was added     
to the resulting image to yield a signal-to-noise ratio similar to that    
of the photometric data we obtained for NGC~772, NGC~3898 and NGC~7782    
during the VATT observing runs.    
    
The images of the artificial galaxies have been analyzed as if they 
were real.  The two-dimensional parametric decomposition has been 
applied using as initial trial parameters the $I_e$, $r_e$, 
$(b/a)_{\it bulge}$, $I_0$, $h$, and $(b/a)_{\it disc}$ values 
obtained with a standard parametric decomposition of the major and 
minor-axis surface brightness profiles measured by fitting ellipses to 
the galaxy isophotes.  The fitting algorithm was able to recover the 
input parameters within the estimated errors with a scatter consistent 
with the results of the similar double-blind tests performed by 
Schombert \& Bothun (1987), Byun \& Freeman (1995) and Wadadekar, 
Robbason \& Kembhavi (1999). 

\begin{table*}   
\caption{Parameters from the dynamical models} 
\begin{center}   
\begin{tabular}{lcccccclccccccc}   
\noalign{\smallskip}   
\hline   
\noalign{\smallskip}   
\multicolumn{1}{c}{Object} &    
\multicolumn{4}{c}{Bulge} &   
\multicolumn{1}{c}{} &    
\multicolumn{4}{c}{Disc} &   
\multicolumn{1}{c}{}  &  
\multicolumn{1}{c}{Dark halo}  &  
\multicolumn{1}{c}{}  &   
\multicolumn{2}{c}{Bulge + Disc} \\   
\noalign{\smallskip}   
\cline{2-5}\cline{7-10}\cline{12-12}\cline{14-15} 
\noalign{\smallskip}  
\multicolumn{1}{c}{} &   
\multicolumn{1}{c}{$L\;^{\rm a}$} &   
\multicolumn{1}{c}{$L\;^{\rm b}$} &   
\multicolumn{1}{c}{$M\;^{\rm c}$} &   
\multicolumn{1}{c}{$M/L$} &   
\multicolumn{1}{c}{}&   
\multicolumn{1}{c}{$L\;^{\rm a}$} &   
\multicolumn{1}{c}{$L\;^{\rm b}$} &   
\multicolumn{1}{c}{$M\;^{\rm c}$} &   
\multicolumn{1}{c}{$M/L$} &   
\multicolumn{1}{c}{}&   
\multicolumn{1}{c}{$M\;^{\rm d}$}  &  
\multicolumn{1}{c}{}  &  
\multicolumn{1}{c}{$M\;^{\rm b}$}   & 
\multicolumn{1}{c}{$M/L$} \\ 
\noalign{\smallskip}   
\multicolumn{1}{c}{} &   
\multicolumn{1}{c}{[\%]} &   
\multicolumn{1}{c}{[\lvsun]} &   
\multicolumn{1}{c}{[\Msun]} &   
\multicolumn{1}{c}{[$\Upsilon_\odot$]} &   
\multicolumn{1}{c}{} &   
\multicolumn{1}{c}{[\%]} &   
\multicolumn{1}{c}{[\lvsun]} &   
\multicolumn{1}{c}{[\Msun]} &   
\multicolumn{1}{c}{[$\Upsilon_\odot$]} &   
\multicolumn{1}{c}{} &   
\multicolumn{1}{c}{[\Msun]} &   
\multicolumn{1}{c}{} &   
\multicolumn{1}{c}{[\Msun]} & 
\multicolumn{1}{c}{[$\Upsilon_\odot$]} \\   
\noalign{\smallskip}   
\noalign{\smallskip}   
\hline  
\noalign{\smallskip}   
NGC~772 & 52\% & $5.8\cdot10^{10}$ & $3.8\cdot10^{11}$ & 6.6 && 48\% & $ 5.4\cdot 10^{10}$ & $1.8\cdot 10^{11}$ & 3.3 &&--&& $5.6 \cdot 10^{11}$&  5.0 \\   
NGC~3898& 67\% & $1.1\cdot10^{10}$ & $7.1\cdot 10^{10}$ & 6.4 && 33\% & $4.8\cdot10^9$ & $2.0\cdot 10^{10}$ & 4.2 && $9.0\cdot10^9$&& $9.1 \cdot 10^{10}$ & 5.8 \\   
NGC~7782& 11\% & $1.03\cdot10^{10}$ & $ 8.5\cdot 10^{10}$ & 8.3 && 89\% & $8.6\cdot 10^{10}$& $ 3.6\cdot 10^{11}$ & 4.2 &&--&& $4.4 \cdot 10^{11}$ & 4.6 \\   
\noalign{\smallskip}   
\hline   
\noalign{\smallskip}    
\label{tab:masses}   
\end{tabular}   
\begin{minipage}{17cm}    
$^{\rm a}$ from the 2D decomposition in Table~\ref{tab:result2Dfit}.\\ 
$^{\rm b}$ adopting the total corrected $V-$band luminosities $L^0_{V_T}$ 
           derived from Table~\ref{tab:properties}. \\ 
$^{\rm c}$ from the dynamical models shown in Figs.~\ref{fig:n772_model},  
           \ref{fig:n3898_model} and \ref{fig:n7782_model}. \\ 
$^{\rm d}$ at the outermost observed radius measured for the  
           ionized-gas component.\\  
\end{minipage}    
\end{center}   
\end{table*}

\subsection{Dynamical models}    
\label{sec:dynmodels}    
    
In order to investigate the gas and stellar kinematics we use the   
self-consistent dynamical models by Pignatelli \& Galletta (1999).  We    
give here just a brief summary of the technique itself and of the    
general assumptions used.    
    
The galaxy can be described by superposition of different    
components.  For each component, we separately assume:    
   
\begin{description}    
\item[-] the density distribution is oblate, without triaxial   
structures;   
\item[-] the isodensity surfaces are similar concentric spheroids;   
\item[-] the surface density profile follows a simple   
$r^{1/4}$ or an exponential law;   
\item[-] the velocity distribution is locally Gaussian;   
\item[-] the velocity dispersion is isotropic   
($\sigma_r = \sigma_\theta =\sigma_z$);   
\item[-] the mass-to-light ratio is constant with radius.   
\end{description}    
    
Our model does not consider the possible presence of triaxial    
structures (bar; triaxial bulge; tilted component; warp) or of    
anisotropy in the velocity distribution (with predominance of radial    
or tangential orbits).    
    
With these assumptions, the model has $4n+1$ free parameters, where 
$n$ is the number of adopted components: namely the luminosity 
$L_{tot}$, scale length ($r_e$ or $h$), mass-to-light ratio $M/L$ 
and flattening $b/a$ of each component plus the inclination angle of 
the galaxy.  In principle, photometry can be used (as explained in 
Sect.~\ref{sec:decomposition}) to constrain all these parameters 
except the mass-to-light ratios and the inclination angle $i$, which must be 
derived by kinematics. 
    
For each given choice of the parameters above, we compute the
gravitational potential of the total mass distribution and integrate
the Jeans equations for the stellar component, obtaining a
self-consistent model of the rotation velocity and
velocity-dispersion profiles which include the asymmetric drift
effects.
       
In order to compare the observed data with the prediction of the    
model, we also need to reproduce the deviations of the LOSVD profiles  
from a pure Gaussian shape. In fact, in    
the regions where the bulge and disc luminosities are comparable we    
expect that the superposition of the rapid rotation of the disc with    
the slower rotation of the other components will produce a clearly   
non-Gaussian, and sometimes even 2-peaked, LOSVD even assuming that    
each individual component has a Gaussian velocity distribution.    
    
We parametrized these deviations in terms of the usual Gauss-Hermite    
series higher moments $h_3$ and $h_4$, that we obtained by means of a    
first-order approximation from the momenta of the model velocity    
distribution (van der Marel \& Franx 1993; Pignatelli \& Galletta    
1999).    
    
Finally, we convolved the results of the model with the seeing and    
took into account the instrumental 
setup used in the different observations (slit    
width, pixel size). The final model profiles are compared with the    
observed stellar kinematic data and the best-fit model is found with    
the help of the standard reduced $\chi^2$-analysis. Masses 
and mass-to-light ratios of different components are given in Table 
\ref{tab:masses}.   
    
Once the fit of the stellar kinematics has been performed, and the 
overall potential of the galaxy is known, one can derive the circular 
velocity $V_c = R (\partial \Phi/ \partial R)$ directly. By 
overlaying the $V_c$ obtained in this way (and corrected for 
inclination) on the observed gas rotation velocity, we can 
immediately notice any deviation from purely circular motion.  The 
evaluation of the different convolution effects is crucial especially 
within the innermost regions, where the rise of the velocity curve 
is smoothed, and the value of the velocity dispersion is increased, by 
the seeing effects.

\section{Results}    
\label{sec:results}    
    
\subsection{NGC 772}    
\label{sec:n772_results}    
    
\subsubsection{Stellar and ionized-gas kinematics}    
\label{sec:n772_kinematics}    
    
\begin{figure}   
\vspace*{13.5cm}    
\includegraphics{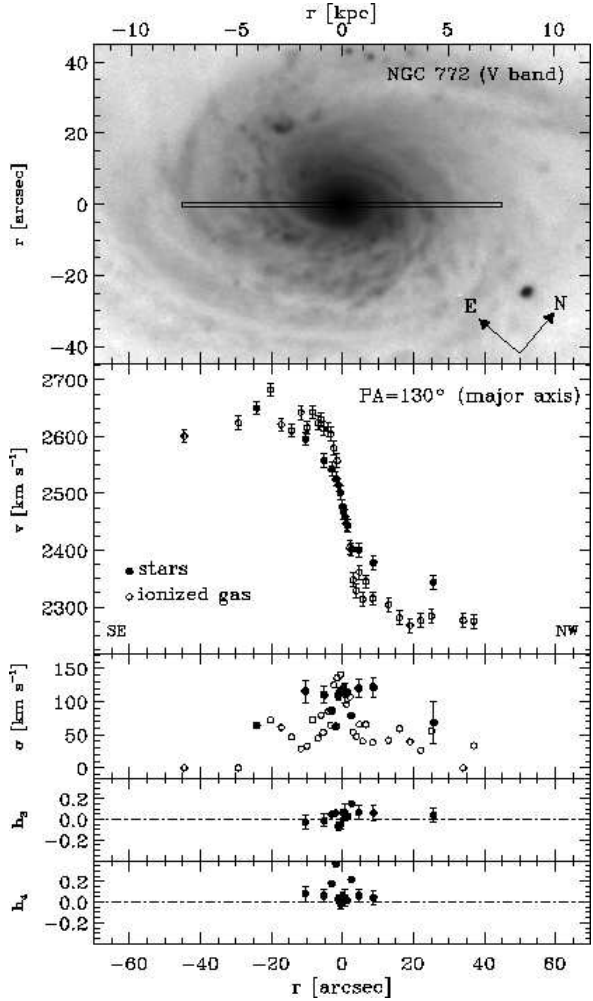}    
\caption[]{The stellar ({\it filled circles\/}) and ionized-gas     
  ({\it open circles\/}) kinematics measured along the major axis     
  ($\rm P.A. = 130^\circ$) of NGC 772.}    
\label{fig:n772_kinematics}    
\end{figure}

The stellar kinematic data extend to more than $30''$ ($5.0$ kpc) on the    
receding side of NGC 772 and to about $24''$ ($4.0$ kpc) on its    
approaching side (Fig. \ref{fig:n772_kinematics}).  In the inner $4''$    
($0.7$ kpc), the rotation velocity of stars increases to $80$    
\kms . At larger radii, it rises more gently out to $120$ \kms\    
at the last measured point.  At the centre the stellar velocity 
dispersion shows a maximum of about $120$ \kms; away from the nucleus 
it falls off to $75$ \kms\ at $4''$. Outside it remains high, 
ranging between $90$ \kms\ and $140$ \kms.

\begin{figure}    
\vspace*{7.5cm}    
\includegraphics{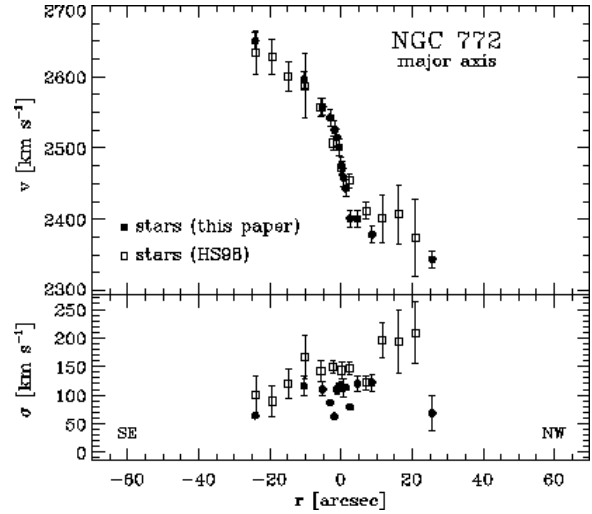}    
\caption[]{NGC 772 major-axis stellar kinematics.  The stellar    
  kinematics derived in the present study at $\rm P.A. = 130^\circ$    
  ({\it filled circles\/}) is shown superimposed on that obtained by    
  Heraudeau \& Simien (1998) at $\rm P.A. = 131^\circ$ ({\it open    
  squares\/}).}     
\label{fig:n772_starcomparison}    
\end{figure}

The stellar velocity curve and velocity-dispersion profile we measured 
along the major axis of NGC 772 are compared in 
Fig. \ref{fig:n772_starcomparison} to those obtained by Heraudeau \& 
Simien (1998).  The two data sets cover almost the same radial region 
and have been collected along two close position angles ($\rm P.A. = 
130^\circ$ and $131^\circ$). Our line-of-sight velocities agree within 
the errors with those measured by Heraudeau \& Simien (1998). However 
their velocity dispersions are higher than those we found.  This is 
only due to the different technique adopted in measuring the stellar 
kinematic parameters rather than to real kinematic features observed 
at the two different position angles.

The ionized-gas velocity is measured out to $45''$ ($7.6$ kpc) on the 
receding side and to less than $40''$ ($6.7$ kpc) on the approaching 
side (Fig. \ref{fig:n772_kinematics}).  The ionized-gas rotation 
velocity has a steeper gradient than the stellar velocity, reaching a 
value of $140$ \kms\ at $|r| \simeq 5''$ ($0.8$ kpc), increasing to 
$175$ \kms\ at $|r| \simeq 10''$ ($1.7$ kpc) and then flattening 
out. The gas velocity dispersion strongly peaks to about $150$ \kms\ 
in the centre; it drops rapidly to values lower than $50$ \kms\ for 
$|r| \ga 5''$.

\subsubsection{$V-$band surface photometry and bulge-disc decomposition}    
\label{sec:n772_photometry}    
    
\begin{figure}    
\vspace*{7cm}    
\includegraphics{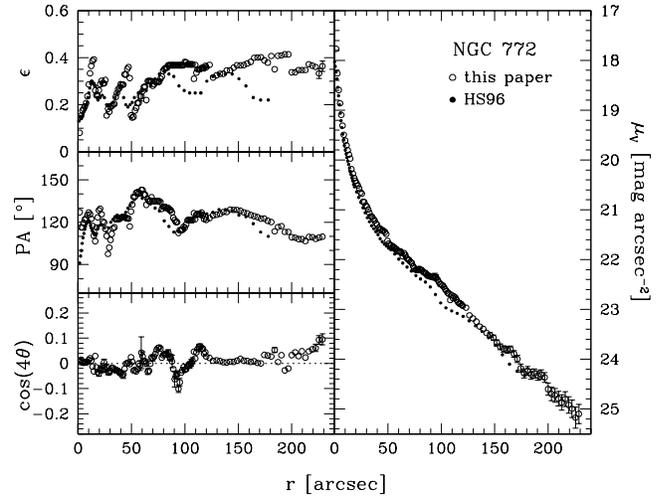}    
\caption[]{The $V-$band surface-brightness, ellipticity, position angle    
  and $\cos{4\theta}$ coefficient profiles we measured
  as a function of the semi-major    
  axis length for NGC 772 ({\it open circles\/}). 
  Error bars smaller than symbols are not plotted.
  The $V-$band surface-brightness, ellipticity and position     
  angle profile from Heraudeau \& Simien (1996, {\it filled circles\/})   
  are also plotted.} 
\label{fig:n772_ellipse}    
\end{figure}

The $V-$band ellipse-averaged radial profiles of surfa\-ce brightness,    
ellipticity, position angle and $\cos{4\theta}$ Fourier
coefficient of NGC 772    
have been measured out to $229''$ ($38.5$ kpc) from the centre    
(Fig. \ref{fig:n772_ellipse}).    
    
\begin{figure}    
\vspace*{7cm}    
\includegraphics{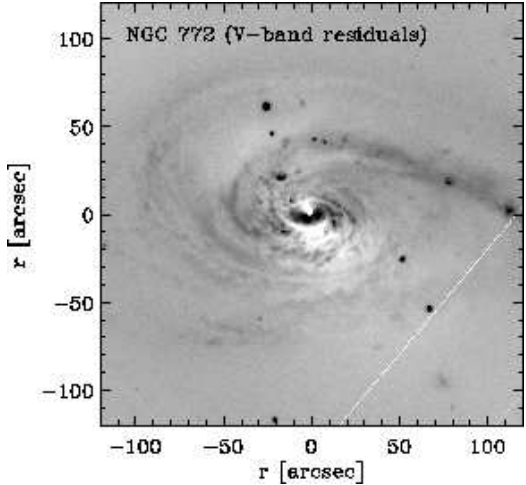}    
\caption[]{The residual image of NGC 772     
  obtained by subtracting the model surface brightness to the observed
  one. The orientation of the image is the same of
  Fig. \protect\ref{fig:n772_kinematics}.  The photometric parameters
  of the model are given in Table \protect\ref{tab:result2Dfit}.}
\label{fig:n772_Vband_residuals}    
\end{figure}

In the inner $60''$ ($10.1$ kpc) the values of ellipticity and 
position angle show a series of bumps and wiggles, oscillating between 
$0.15$ and $0.36$ and between $98^\circ$ and $143^\circ$, 
respectively.  These features are due to the inner spiral arms of the 
galaxy.  The bump in the surface-brightness profile at $90''$, and the 
corresponding abrupt variations in the position angle and fourth-order 
cosine Fourier coefficient can be ascribed to the prominent northern 
arm (see Fig.~\ref{fig:n772_Vband_residuals}). At larger radii, the 
galaxy light is dominated by the contribution of the disc component 
characterized by an exponential surface-brightness profile and almost 
constant ellipticity ($\epsilon \simeq 0.35$) and position angle ($\rm 
P.A. \simeq 120^\circ$). 
    
In Fig. \ref{fig:n772_ellipse} our data are compared to those 
obtained by Heraudeau \& Simien (1996) in the same band. The 
photometric profiles of Heraudeau \& Simien (1996) extend out 
to $180''$ from the centre and 
their surface brightness is $\la0.4$ \mas\ fainter than our.
In the inner $50''$ their position angle and ellipticity 
data have smoother radial trends than those we measured. We noticed 
that the position-angle values included by Heraudeau \& Simien (1996) 
in Table 5 of their paper differ from those they plotted in their 
Fig. 10.  In our comparison we adopted the plotted ones. The 
disagreement between the two sets of data can be explained if the 
tabulated values result from an isophotal ellipse fitting performed on 
an image of NGC 772 with an incorrect orientation. 
       
In Fig. \ref{fig:n772_Vband_residuals} we show the result of the 
bulge-disc photometric decomposition of the surface-brightness 
distribution of NGC 772, which has been performed with the parametric 
two-dimensional technique discussed in Sec. \ref{sec:decomposition}. 
The spiral arms extending into the very centre of NGC 772 are clearly 
visible in the residual image obtained by subtracting the model 
surface brightness of the galaxy from the observed surface 
brightness. These structures resemble those detected by Carollo et 
al. (1997, 1998) in several Sa -- Sbc galaxies, in which the spiral 
pattern is visible down to the innermost radius accessible to the HST 
WFPC2 imaging.

By fitting an exponential profile to the ellipse-averaged    
surface-brightness profile of NGC 772 measured in the $B$ band, Lu    
(1998) derived the following disc parameters $\mu_0 = 22.01$ \mas, $h    
= 72\farcs4$ and $b/a = 0.58$. Our exponential disc results    
larger ($h = 96\farcs3$) and slightly more inclined    
($b/a=0.39$) than that obtained by Lu (1998).  This is due to the    
different decomposition technique we adopted more than to the different    
bandpass in which we observed. In fact Lu (1998) fitted his   
exponential disc directly to the observed surface-brightness profile in the   
radial range between $75\farcs1$ and $133\farcs1$ (which he  
judged by eye to be disc-dominated) without taking into account   
any contribution from the conspicuous bulge component.

\subsubsection{Ionized-gas distribution}    
\label{sec:n772_Halpha}    
    
\begin{figure}    
\vspace*{14.5cm}    
\includegraphics{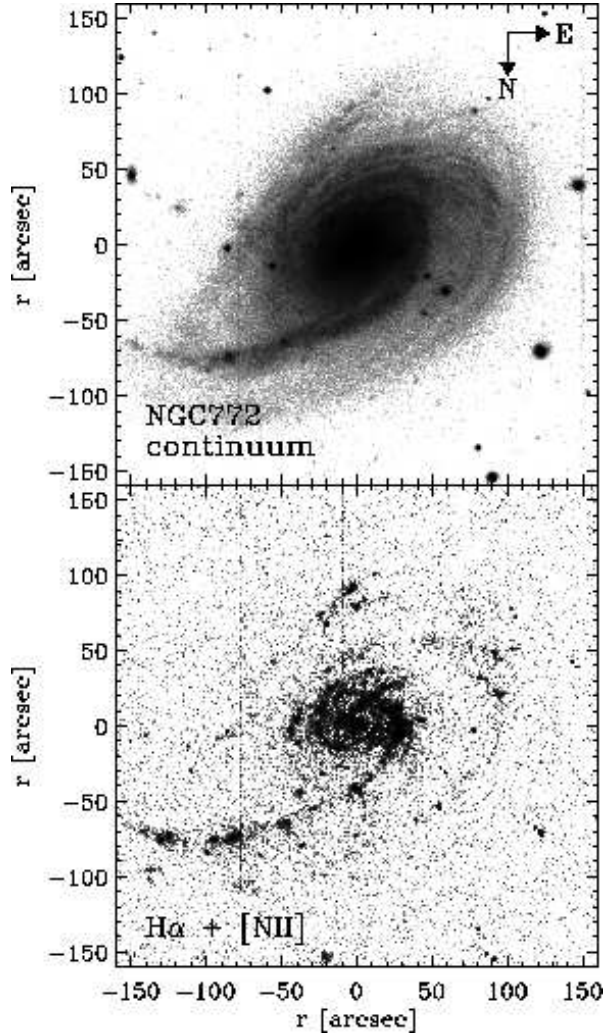}    
\caption[]{The stellar-continuum image ({\it upper panel\/}) and the    
  continuum-subtracted \ha$+$\nii\ emission image  ({\it lower panel\/})    
  of NGC 772.}    
\label{fig:n772_Halpha}    
\end{figure}

In NGC 772 most of the \hii\ regions visible in our    
\ha$+$\nii\ image (Fig. \ref{fig:n772_Halpha})    
lie along the two arms, extending to the northern side of the    
galaxy, and along the short and double-ended arm opposite to them.

The ionized-gas emission is concentrated in the region of the galaxy 
($r \la 40''$) bounded by the pseudo-ring formed by the two 
tightly-wound northern spiral arms. In the \ha$+$\nii\ emission map it 
is possible to disentangle an outer region extending between $10''$ 
and $40''$ and characterized by smooth and diffuse emission, from the 
inner region ($r \la 10''$) characterized by more 
concentrated emission and contributing about $5\%$ of the total 
flux. 
    
The ionized gas between $10''$ and $40''$ shows the kinematic    
behaviour typical of a disc ($V/\sigma \simeq 5$) as inferred from its    
high rotation velocity and low velocity dispersion    
(Fig. \ref{fig:n772_kinematics}).    
    
\begin{figure*}    
\vspace*{19.5cm}    
\includegraphics{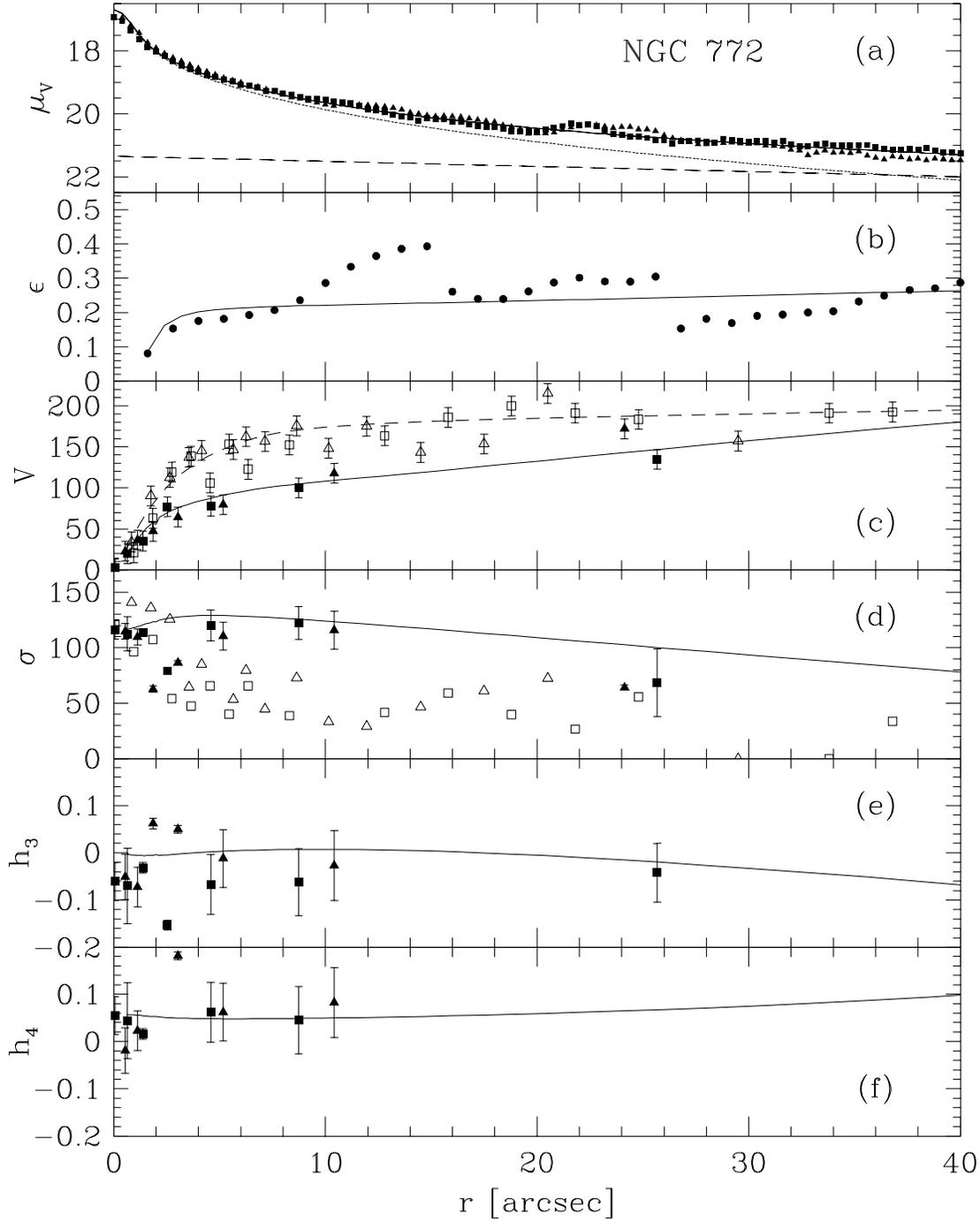}    
\caption[]{Photometric and dynamical radial profiles for NGC~772 with their    
  respective best-fit curves obtained from our models.     
  {\bf (a):} $V-$band surface brightness ({\it filled   
  symbols\/}) as a function of radius along the major axis. We also   
  plot the model bulge ({\it dotted line\/}) and disc ({\it   
  dashed lines\/}) surface-brightness profiles, together with their   
  sum convolved with the seeing ({\it solid line\/}).    
  {\bf(b):} Ellipse-averaged ellipticity for the $V$-band data    
  ({\it filled circles\/}) and the photometric model ({\it solid line\/}).     
  {\bf (c):} Observed stellar velocity ({\it filled symbols\/})   
  with its associated model ({\it solid line\/}), and the   
  ionized-gas velocity ({\it open symbols\/}). The   
  {\it dashed line\/} is the circular velocity inferred from the   
  dynamical model.   
  {\bf (d):} the same as (c) but for the velocity dispersion.      
  {\bf (e):} the same as (c) but for $h_3$ coefficients of the   
  Gauss-Hermite expansion of the line profile of the stars.     
  {\bf (f):} the same as (c) but for $h_4$ coefficients of the   
  Gauss-Hermite expansion of the line profile of the stars.   
  In all the plots the {\it square symbols\/} and the   
  {\it triangle symbols\/} represent data derived for the approaching   
  NW side and for the receding SE side, respectively.}    
\label{fig:n772_model}    
\end{figure*}

In the inner region the kinematic behaviour of the gas is   
different. The observed gas velocity dispersion (ranging between $70$   
and $150$ \kms) is comparable to that of the stellar component and   
therefore is far higher than that expected from thermal motion or   
small-scale turbulence ($\sigma \la 50$ \kms) typically observed in a   
gaseous disc.  This may well be an indication that in this region the gas   
is being dynamically supported by pressure (i.e. random motions)    
rather than rotation, as it occurs in the outer part of the galaxy.   
There is, however, a wide region ($4''<r<10''$) in which the gas   
velocity dispersion is still only marginally greater than 50 \kms, and   
appears to be lower than its stellar counterpart   
(see panel (b) in Fig. \ref{fig:n772_model}). The high    
velocity dispersions observed near the nucleus could then only be an   
observational effect due to seeing. In order to discriminate between   
the two cases, the application of a dynamical model is required.

\subsubsection{Dynamical modelling}    
\label{sec:n772_modelling}    
  
In Fig. \ref{fig:n772_model} we show for NGC 772 the 
comparison between the photometric and kinematic data 
and the best-fit photometric dynamical models we obtained 
with the techniques discussed in 
Sec.~\ref{sec:structure}. The mass-to-light ratios, 
flattening and masses of bulge and disc stellar 
components of NGC 772 derived with the dynamical modelling 
described in Sec. \ref{sec:dynmodels} are given in  
Table~\ref{tab:masses}. 
         
The agreement between the stellar kinematic data and the 
predictions of the dynamical model are fairly good at all 
radii, apart from the features observed in the LOSVD at 
$|r|\simeq3''$. At this radius we notice a sharp decrease 
(from 115 \kms\ to 80 \kms) in the observed velocity 
dispersion and very strong asymmetries in the line shape, 
while the rotation velocity does not show any 
significant variation. The model is unable to reproduce 
these abrupt but radially confined changes in the 
velocity-dispersion, $h_3$ and $h_4$ profiles.  Either 
absorbing dust or nuclear peculiar motions could be 
responsible for this feature. 
   
It is worth noting that the model reproduces the observed 
gaseous and stellar rotation velocity without the need 
of introducing a dark halo. According to the photometric 
decomposition, our kinematic data are limited to the 
bulge-dominated region of NGC 772 where the luminous mass 
is expected to dominate the galaxy dynamics. In agreement 
with these results, the dynamical model shows that 
asymmetric drift has a sizable effect on the stellar 
component in the whole observed region as it can be 
desumed by considering that we derived a difference of 35 
\kms\ between stellar rotation velocity and circular 
speed at the farthest measured point. 
  
Once the convolution with seeing and instrumental setup 
has been taken into account, the gas appears to rotate in 
almost circular orbits.  We conclude that the gas 
rotation curve of this galaxy can not be classified as a 
slowly-rising one, 
and that the observed central rise in the gas velocity 
dispersion is due to seeing effects. 
 
\subsection{NGC 3898}    
\label{sec:n3898_results}    
    
\subsubsection{Stellar and ionized-gas kinematics}    
\label{sec:n3898_kinematics}    
    
\begin{figure}    
\vspace*{13cm}    
\includegraphics{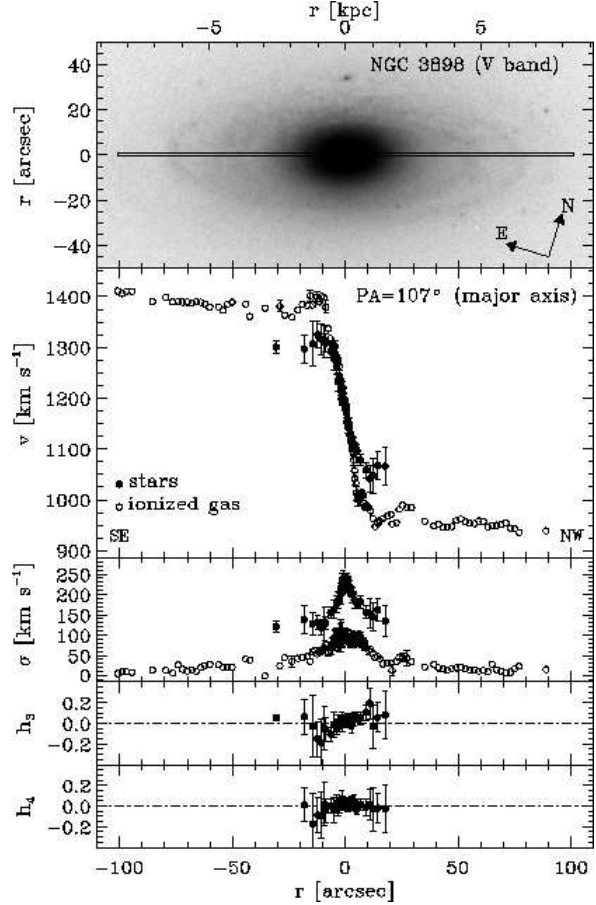}    
\caption[]{The stellar ({\it filled circles\/}) and ionized-gas     
  ({\it open circles\/}) kinematic parameters measured along the major axis     
  ($\rm P.A. = 107^\circ$) of NGC 3898.}    
\label{fig:n3898_kinematics}    
\end{figure}

\begin{figure}    
\vspace*{7.5cm}    
\includegraphics{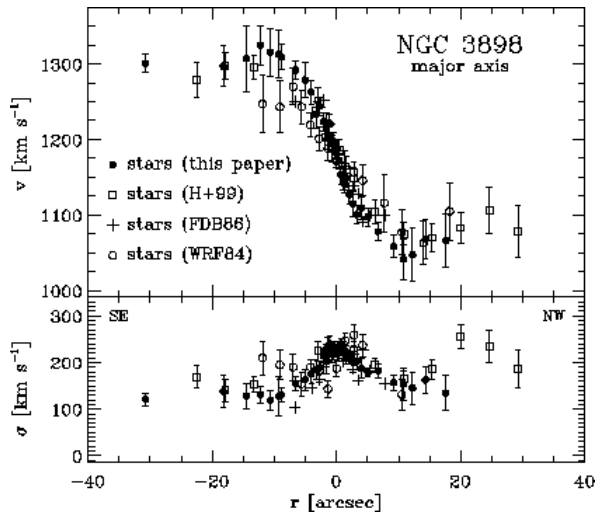}    
\caption[]{NGC 3898 major-axis stellar kinematic parameters.  The stellar    
  velocity and velocity dispersion derived in the present study  
  at $\rm P.A.=107^\circ$ ({\it filled circles\/}) are    
  shown superimposed on those obtained by Whitmore et al. (1984, {\it    
  open circles\/}), Fillmore et al. (1986, {\it crosses\/}), and    
  Heraudeau et al. (1999, {\it open squares\/}) for the same    
  position angle.}    
\label{fig:n3898_starcomparison}    
\end{figure}

The stellar kinematic parameters are observed out to $20''$ ($1.7$
kpc) and $30''$ ($3.4$ kpc) on the approaching and receding side of
NGC 3898, respectively (Fig. \ref{fig:n3898_kinematics}). The stellar
rotation velocity increases linearly up to $130$ \kms\ in the inner
$10''$ ($0.8$ kpc), then it decreases to $90$ \kms\ at the last
observed radius. At the centre, the stellar velocity dispersion is 
about 220 \kms, then increases up to a maximum value of $240$ \kms\ at 
$|r|\simeq 1''$ (Fig. \ref{fig:n3898_starcomparison}), suggesting the 
presence of an unresolved kinematically decoupled core component (see 
Friedli 1996 for a discussion about this feature in the
velcity-dispersion profile).  Off the nucleus, 
the stellar velocity dispersion decreases to $160$ \kms\ on the 
approaching side and $130$ \kms\ on the receding side. 
    
The stellar velocity curve and velocity-dispersion profile measured    
along the major axis of NGC 3898 are compared  
in Fig. \ref{fig:n3898_starcomparison} to those obtained by    
Whitmore et al. (1984), Fillmore et al. (1986)    
and by Heraudeau et al. (1999).    
Although the velocity data agree within the errors, a difference in    
the central velocity gradients measured by the various groups is present.    
In particular, there is a velocity discrepancy    
(ranging between $100$ and $150$ \kms) between our    
outer rotation velocities and those by Whitmore et al. (1984),   
who also measured a higher velocity dispersion on the    
receding side.

The ionized-gas kinematic parameters extend to $90''$ ($7.5$ kpc) on 
each side of the nucleus (Fig. \ref{fig:n3898_kinematics}).  The gas 
rotation velocity has a steeper gradient than the stellar velocity, 
reaching $200$ \kms\ at $|r| \simeq 10''$ and remaining almost 
constant at this value further out.  The gas velocity dispersion is 
about $90$ \kms\ for $|r| \la 10''$ decreasing at larger radii ($|r| 
\ga 30''$) to $20$ \kms\ on both sides of the nucleus. 
    
\begin{figure}    
\vspace*{7.5cm}    
\includegraphics{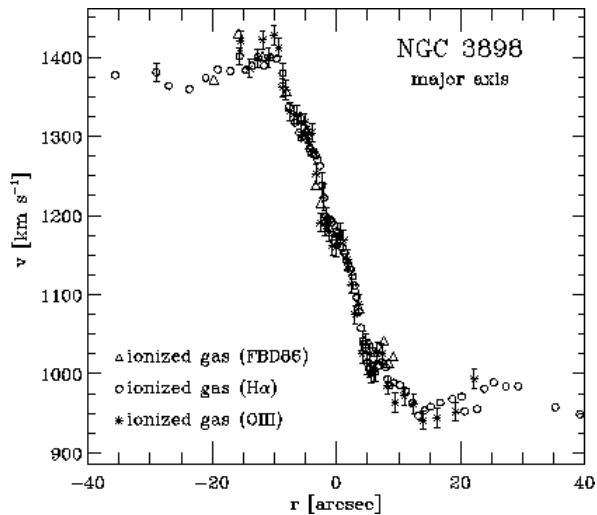}    
\caption[]{NGC 3898 major-axis ionized-gas kinematic parameters.  The    
  ionized-gas velocities derived in the present study at $\rm P.A. =    
  107^\circ$ using the spectra obtained at INT ({\it open circles\/})   
  and MMT ({\it asterisks\/}) are shown superimposed on those    
  obtained measuring the \ha\ emission line by Rubin et al. (1985, {\it    
  crosses\/}) and Fillmore et al. (1986, {\it open triangles\/})    
  along the same position angle.}    
\label{fig:n3898_gascomparison}    
\end{figure}

The ionized-gas rotation curves we measured along the major axis of 
NGC 3898 from the spectra obtained with INT and MMT are compared in 
Fig. \ref{fig:n3898_gascomparison} to the \ha\ rotation velocities 
obtained by Rubin et al. (1985) and Fillmore et al. (1986).  There is 
a good agreement between the line-of-sight velocities obtained in the 
different runs. In particular, all the rotation curves closely match 
each other, showing the same wiggles and bumps in the velocity for 
$|r|<8''$. However, between $8''$ and $20''$ on both sides of the 
nucleus Rubin et al. (1985) measured a slower rotation than either we 
or Fillmore et al. (1986) derived. As far as the gas velocity 
dispersion is concerned, only our own data are available. They are 
based on the line width of the 
\oiii\ and \ha\ emission lines present in the MMT and INT spectra.  
The two data set are consistent within the scatter of the points 
derived from the \oiii\ line.

\subsubsection{$V-$band surface photometry and bulge-disc decomposition}    
\label{sec:n3898_photometry}    
    
\begin{figure}    
\vspace*{7cm}    
\includegraphics{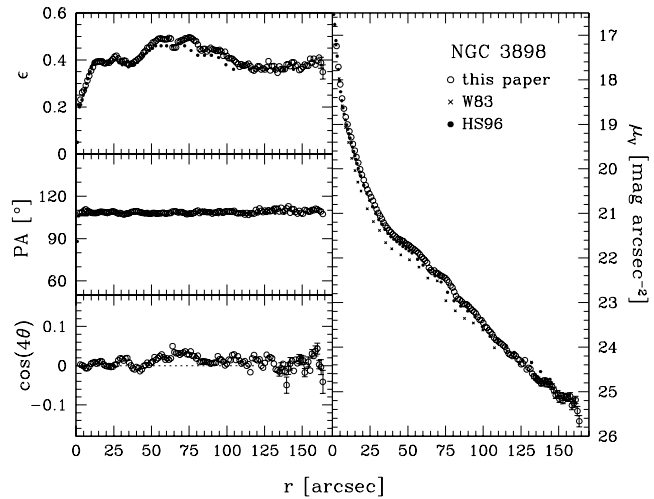}    
\caption[]{The $V-$band surface-brightness, ellipticity, position angle    
  and $\cos{4\theta}$ coefficient profiles we measured
  as function of the semi-major axis length for NGC 3898
  ({\it open circles\/}). Error bars smaller than symbols are not plotted.
  The $V-$band surface-brightness, ellipticity and position     
  angle profiles from Heraudeau \& Simien (1996, {\it filled circles\/}) 
  and the surface-brightness profile by Watanabe (1983,    
  {\it crosses\/}) are also shown.} 
\label{fig:n3898_ellipse}    
\end{figure}

The $V-$band radial profiles of surface brightness, ellipticity,    
position angle and $\cos{4\theta}$ Fourier coefficient measured by fitting    
ellipses to the NGC 3898 isophotes extend out to $164''$ ($13.6$ kpc)    
from the centre (Fig. \ref{fig:n3898_ellipse}).

The ellipticity rises from $0.23$ to $0.39$ in the inner $12''$    
($1.0$ kpc) remaining almost constant out to $40''$ ($3.3$ kpc).    
Between $40''$ and $110''$ ($9.1$ kpc) it increases to a maximum of    
$0.5$ at $75''$ ($6.2$ kpc) and decreases to the initial value of    
0.39. Further out it remains almost constant at $0.37$ out to the    
farthest observed radius. The isophotal position angle is constant at    
$109^\circ$ at all radii. The isophotes are slightly disky between    
$40''$ and $85''$ ($7.1$ kpc) and elliptical elsewhere. The    
surface-brightness profile is extremely regular with a few small bumps    
due to the spiral arms.    
    
In Fig. \ref{fig:n3898_ellipse} our surface-brightness radial  
profile is compared to those obtained by Watanabe (1983), and  
Heraudeau \& Simien (1996), who also measured for NGC 3898 the  
ellipticity and position angle radial profiles.  The agreement between  
our data and those of Heraudeau \& Simien (1996) is good. The surface  
brightness measured by Watanabe (1983) in the radial region between   
$10''$ and $100''$ is $\la0.5$ \mas\ fainter than ours.  The  
surface-brightness profile of Kodaira et al. (1990) is not shown in  
Fig. \ref{fig:n3898_ellipse}. In their Photometric  
Atlas of Northern Bright Galaxies only the profiles extracted along  
the major and minor axes of NGC 3898 are available and there is not
an ellipse-averaged profile. This is also true for NGC 7782.  
          
\begin{figure}    
\vspace*{7cm}    
\includegraphics{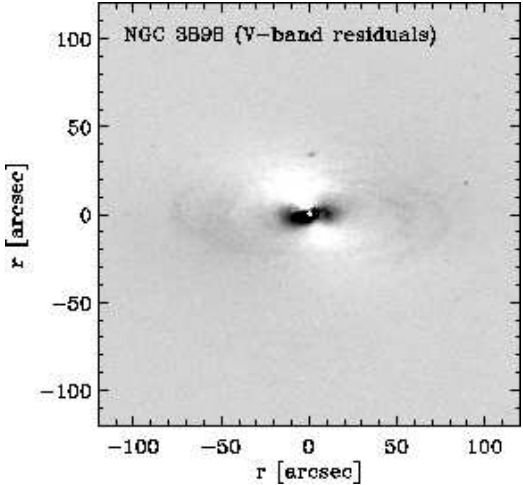}    
\caption[]{The residual image of NGC 3898     
  obtained by subtracting the model surface brightness to the observed
  one. The orientation of the image is the same of
  Fig. \protect\ref{fig:n3898_kinematics}.  The photometric parameters
  of the model are given in Table \protect\ref{tab:result2Dfit}.}
\label{fig:n3898_Vband_residuals}    
\end{figure}

The result of the two-dimensional bulge-disc parametric 
decomposition of the surface-brightness distribution of 
NGC 3898 is shown in Fig. \ref{fig:n3898_Vband_residuals}.  
The spiral pattern shows up in the panel of the 
residuals, as well as a central structure due to the 
difference in this region between the observed 
ellipticity (increasing from $0.28$ to $0.39$) and the 
constant value ($\epsilon_{\it bulge} = 0.39$) derived 
for the bulge component.

The surface-brightness profiles of Watanabe (1983) and Kodaira et al.  
(1990) were fitted with a $r^{1/4}$ bulge and an exponential disc by van  
Driel \& van Woerden (1994) and Baggett et al. (1998),  
respectively.  Van Driel \& van Woerden (1994) using Burstein's (1979)  
method obtained the following photometric parameters $\mu_e = 17.1$ \mas,  
$r_e=4\farcs4$, $\mu_0=20.6$ \mas, and $h=33\farcs0$. The best-fit  
parameters derived by Baggett et al. (1998), performing an  
intensity-weighted fit between $3''$ and $156''$ to avoid the portion of  
the profile most affected by seeing, are $\mu_e = 18.3$ \mas,  
$r_e=6\farcs2$, $\mu_0=20.8$ \mas, and $h=37\farcs8$. The results of van  
Driel \& van Woerden (1994) and Baggett et al. (1998) are consistent if we  
take into account that they used an ellipse-averaged profile and the  
profile extracted along the galaxy major axis, respectively.   
The bulge and disc scale parameters we obtained for NGC 3898 are   
quite different since we have taken properly into account the  
apparent axial ratios of the two components.  
      
\subsubsection{Ionized-gas distribution}    
\label{sec:n3898_Halpha}    
    
\begin{figure}   
\vspace*{14.5cm}    
\includegraphics{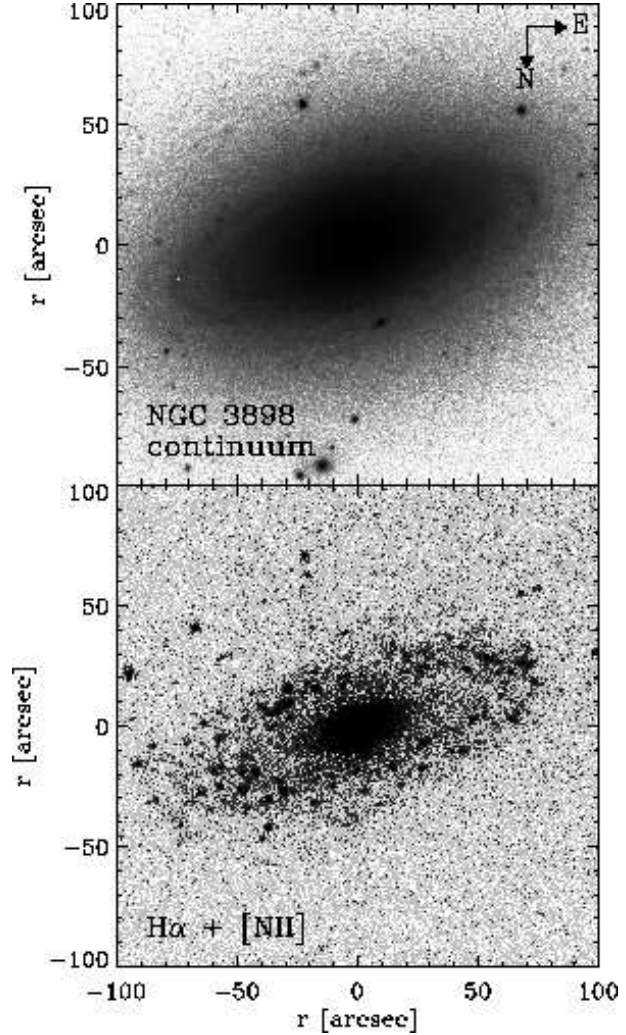}    
\caption[]{The stellar-continuum image ({\it upper panel\/}) and the    
  continuum-subtracted \ha$+$\nii\ emission image  ({\it lower panel\/})    
  of NGC 3898.}    
\label{fig:n3898_Halpha}    
\end{figure}

Ho, Filippenko \& Sargent (1997) classified the nucleus of NGC 3898 as 
intermediate between an \hii\ nucleus and a LINER.  This bright 
nucleus is also visible in our \ha$+$\nii\ emission image of NGC 3898 
(Fig. \ref{fig:n3898_Halpha}), mapping the ionized-gas distribution of 
the galaxy. The ionized-gas emission is smooth and featureless in the 
bulge-dominated region ($r \la 20''$), where about $10\%$ of the warm 
gas resides.  It shows a clumpy and fragmented distribution in the 
disc-dominated region where numerous \hii\ regions trace the NGC 3898 
multiple-arm structure. The transition between the two regions occurs 
abruptly at $35''$ from the centre where the gas emission in the 
major-axis spectrum (Fig. \ref{fig:n3898_kinematics}) cuts off. The 
distribution of the \hii\ regions (which closely follows the 
continuum isophotes) indicates that the gas and stellar discs are 
coplanar. 
 
\begin{figure}    
\vspace*{8cm}    
\includegraphics{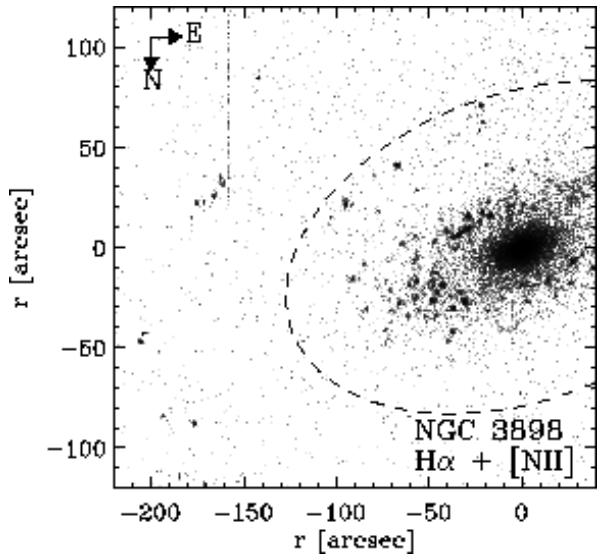}    
\caption[]{Continuum-subtracted \ha$+$\nii\ emission image     
  of NGC 3898 showing the \hii\ regions tracing the two external faint    
  arms of the galaxy. The {\it dashed line\/} indicates the optical    
  size of the NGC 3898 corresponding to the $D_{25}$ $B-$band diameter
  given in Table \protect\ref{tab:properties}.}    
\label{fig:n3898_arms}    
\end{figure}

On the SW side of the galaxy two series of emission knots
are aligned at a projected angular distance of $80''$ and
$180''$ from the centre, respectively
(Fig. \ref{fig:n3898_arms}). These \hii\ regions trace
the two external faint arms located far beyond the galaxy
main body as discussed in the Carnegie Atlas of Galaxies
(Sandage \& Bedke, 1994, Panel 79).
    
\subsubsection{Dynamical modelling}    
\label{sec:n3898_modelling}    
 
\begin{figure*}  
\vspace*{21.5cm}  
\includegraphics{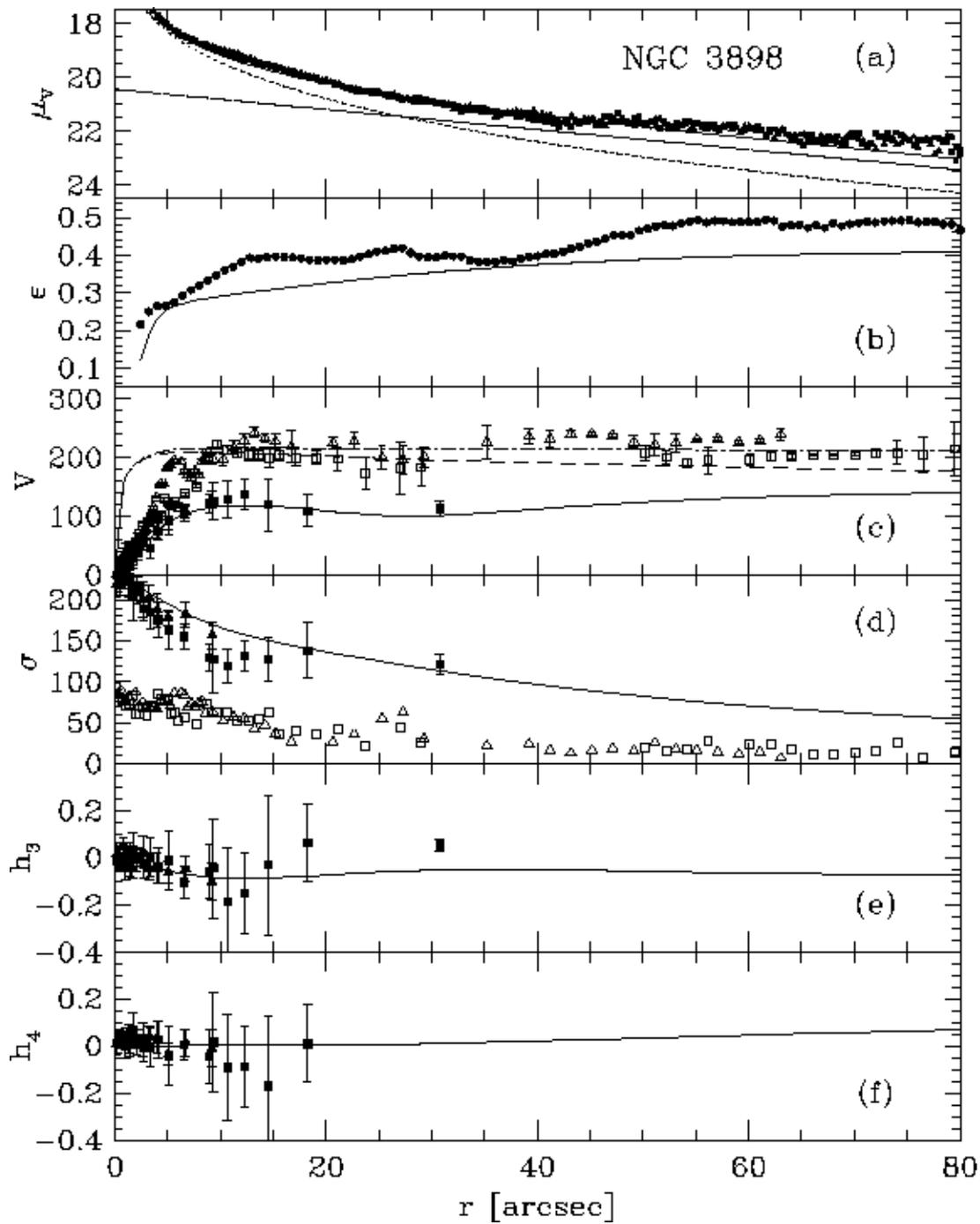}  
\caption[]{Same as Fig.~\protect\ref{fig:n772_model} but for NGC 3898.  
 The {\it squares\/} and {\it triangles\/} represent data measured on 
 the  receding SE and approaching NW sides, respectively. In panel (c)  
 the model circular velocity is plotted with ({\em dot-dashed line}) and  
 without ({\em dashed line}) the contribution of a dark matter halo with  
 the structural parameters given in Table \protect\ref{tab:masses}. 
}  
\label{fig:n3898_model}  
\end{figure*}

In Fig. \ref{fig:n3898_model} we show the comparison between the  
photometric and kinematic data and the best-fit dynamical model  
obtained with the technique discussed in Sec.~\ref{sec:dynmodels}.  
The mass-to-light ratios of the bulge and disc stellar component are shown  
in Table \ref{tab:masses}.  
       
A number of kinematic features highlight the presence of a 
hot spheroidal component.  At $|r|=12''$ both $h_3$ and $h_4$ 
show the presence of a non Gaussian LOSVD which can be 
interpreted as due to the superposition of a radiply 
rotating disk and an almost non-rotating bulge.  The 
stellar rotation velocity is everywhere less than 2/3 
of the gas velocity, as a result of the asymmetric drift 
effect.  Finally, for $12'' \la |r| \la 25''$ the stellar 
velocity dispersion shows a plateau at a value of 110 
\kms, which is the signature of a thick, dynamically hot 
component. 
 
The gas rotation velocity is well approximated by the 
circular velocity of our models for $|r|>8''$. Out to 
$80''$ from the centre the gas rotation curve does not 
show hints of an asymmetric drift effect and is almost flat 
at the constant value of about $200$ \kms.  Kent (1988) 
and Moriondo et al. (1998b) already pointed out that 
for $|r|\la8''$ the gas rotates more slowly than expected 
on the basis of the stellar kinematics and of the 
photometry.  The comparison of the observed gas 
kinematics with our self-consistent dynamical model 
confirms this result. Moreover, we can also conclude that 
seeing convolution and finite slit width can not account 
for this effect, since these parameters were already 
included in the computation of our modeled velocity 
curves.  We stress that, for this particular object, the 
gas rotation curve can not be interpreted as a sign of 
`pressure-supported' gas in the central regions of the 
galaxy. First, the observed gas velocity dispersion is 
too low to account for the large differences between the 
circular velocity computed from the model and the 
observed gas rotation velocity.  Moreover, the gas 
kinematics show strong asymmetries in the region where 
the `slowly rising' effect takes place, with a difference 
of more than 80 \kms\ in the rotation velocities of the 
leading and receding sides. Such asymmetries are not seen 
in the gas velocity-dispersion profile. It is likely 
that other effects are at work here; however, the 
available data do not allow us to distinguish between 
non-axisymmetric distortion of a gaseous disc and 
possible non-equilibrium motions of the gas. 
 
For this galaxy only we were not able to reproduce the 
gas rotation velocities at large radii without adding a 
dark matter halo. For sake of simplicity, we adopted the 
`maximum bulge+disc' paradigm, in order to minimize the 
amount of dark matter needed to fit the data.  A more 
precise measurement of the mass and distribution of the 
dark matter would require a best-fitting procedure of the 
observed kinematic curves varying simultaneously all the 
parameters involved, but can not be applied to this 
object because of the limited extension of the gas 
rotation curve ($\approx 2 h$; see Persic, Salucci \& 
Stel 1996 for a discussion on the subject).  We assumed a 
pseudo-isothermal profile for the dark halo with a 
circular velocity, given by: 
 
\begin{equation}  
V_c^2(r) =v_h^2\left[1-{r_h \over r}\arctan\left({r \over 
r_h}\right)\right]  
\label{eq:vc_pseusoiso}  
\end{equation}  
 
Due to the limitations of the available data, we were 
unable to determine both the total mass and scale radius 
of the halo. Nevertheless, the total mass {\em within} 
at the at the outermost observed radius was quite well 
constrained and turned out to be $9\cdot 10^9$ \Msun, 
with an error of less than $10\%$.  We also found that 
the scale length of the dark halo is so large ($r_h > 
65''$), that in the observed radial range the circular 
velocity contributed by the halo rises almost linearly 
with radius. 
 
Total mass estimates and mass decompositions for this 
galaxy are already present in literature (Fillmore et 
al. 1986; Kent 1998; Moriondo et al. 1998b).  Remarkably, 
despite the various different hypotheses adopted, our 
result for the total luminous mass agrees within $20\%$ 
with all these authors: when scaled to our adopted 
distance of $D=17.1$ Mpc, Kent (1988) gives $M_{\it lum}= 
1.1\cdot 10^{11}$ \Msun, Fillmore et al. (1986) $M_{\it 
lum}= 1.1\cdot 10^{11} $ \Msun, and the most recent 
estimate, given by Moriondo et al. (1998b) is $M_{\it 
lum}= 9.2\cdot 10^{10} $ \Msun, almost identical to our 
value of $M_{\rm lum}= 9.1\cdot 10^{10}$ \Msun. 
On the contrary, the mass decomposition of the different 
authors differs from each other by large factors, being 
strongly dependent on the hypothesis adopted.  Kent 
(1988) and Moriondo et al. (1998b) derive bulge masses, 
which are respectively 15 and 3 times lower than the value 
we obtained. This is due to their assumption that 
the gas rotation velocities were tracing the potential 
everywhere. As a consequence unphysical values for the 
mass-to-light ratios of both components are found, as the authors 
correctly remarked.  This underestimate of the bulge 
contribution to the total mass content leads also to 
an overestimate the disc mass (by a factor of 4, if compared 
with our results), and the mass of the dark halo (in the 
case of Moriondo et al. 1998b). 
 
Fillmore et al. (1986) derived the mass-to-light ratios from the 
stellar kinematics rather than from the gas; however, 
their photometric decomposition is one-dimensional and 
leads to a smaller and more compact bulge in combination 
with a larger disc than the ones we derived.  Therefore 
their results are not easily comparable with ours.  On 
the other hand, our two-dimensional photometric 
decomposition give a bulge-to-disc ratio in good 
agreement with both Kent (1988) and Moriondo et 
al. (1998b), who did not use for their decomposition 
the ellipse-averaged surface-brightness profiles as done 
by Fillmore et al. (1986). 
 
\subsection{NGC 7782}    
\label{sec:n7782_results}    
    
\subsubsection{Stellar and ionized-gas kinematics}    
\label{sec:n7782_kinematics}    
    
\begin{figure}    
\vspace*{12.5cm}    
\includegraphics{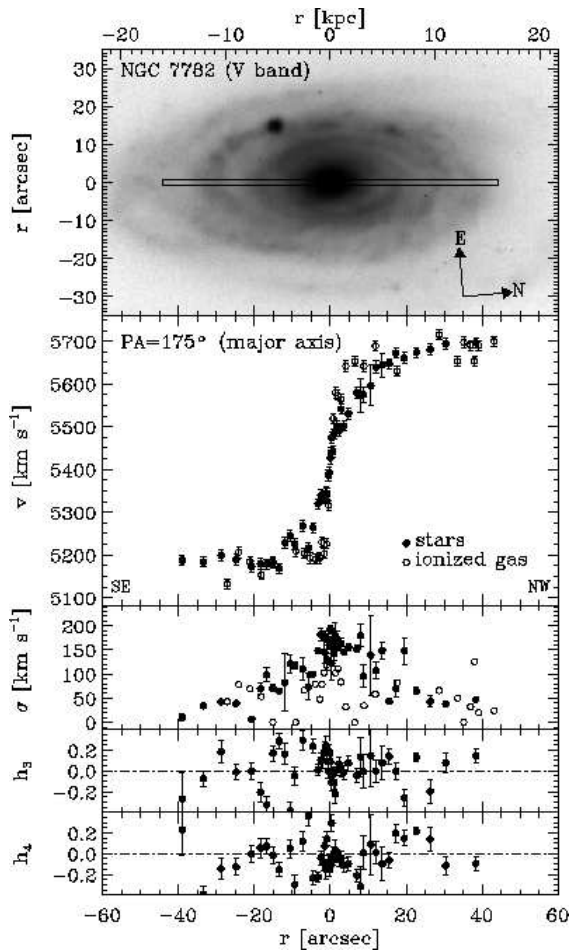}    
\caption[]{The stellar ({\it filled circles\/}) and ionized-gas     
  ({\it open circles\/}) kinematic parameters measured along the major axis     
  ($\rm P.A. = 175^\circ$) of NGC 7782.}    
\label{fig:n7782_kinematics}    
\end{figure}

The stellar velocity curve is observed out to $40''$ ($14.6$ kpc) on 
both sides of NGC 7782 (Fig. \ref{fig:n7782_kinematics}).  The stellar 
rotation velocity increases linearly up to $70$ \kms\ in the inner 
$1''$ ($0.4$ kpc). It does not change for $1'' \la |r| \la 3''$ ($1.1$ 
kpc), and further out it shows a shallower gradient rising to $240$ 
\kms\ at $14''$ ($5.1$ kpc). Outwards the stellar rotation velocity 
remains constant or possibly even rises.  At the centre the stellar 
velocity dispersion peaks to about $190$ \kms, then off the nucleus it 
decreases gradually to less than $50$ \kms\ at the last outermost 
radii. 
    
The ionized-gas velocity curve extends to less than $30''$ ($10.9$ 
kpc) on the approaching side and to less than $45''$ ($16.4$ kpc) on 
the receding side of the galaxy (Fig. \ref{fig:n7782_kinematics}). The 
gas rotation velocity has a steeper gradient than the stellar curve, 
rising to $240$ \kms\ at $3''$; then it remains almost constant, 
becoming similar to the stellar velocity for $|r| > 14''$. The gas 
velocity dispersion is centrally peaked at about $180$ \kms, dropping 
to values lower than $100$ \kms\ for radii larger than $3''$ on both 
sides of the galaxy. Further out it oscillates around values of about $40$ 
\kms. The gas velocity dispersion is not greater than the stellar dispersion
over the whole observed radial range. 
    
\subsubsection{$V-$band surface photometry and bulge-disc decomposition}    
\label{sec:n7782_photometry}    
    
\begin{figure}    
\vspace*{7cm}    
\includegraphics{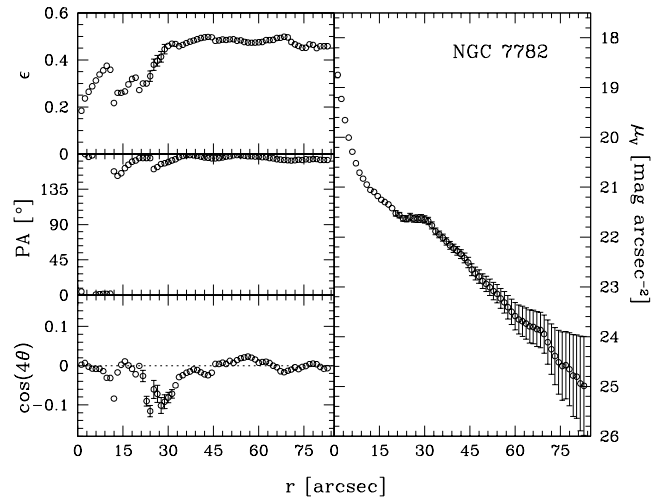}    
\caption[]{The $V-$band surface-brightness, ellipticity, position angle    
  and $\cos{4\theta}$ coefficient profiles as function of the semi-major    
  axis length for NGC 7782. Error bars smaller than symbols are not plotted.}
\label{fig:n7782_ellipse}    
\end{figure}

The $V-$band radial profiles of surface brightness, ellipticity,    
position angle and $\cos{4\theta}$ Fourier coefficient of NGC 7782 are    
measured out to $83''$ ($30.3$ kpc) from the centre    
(Fig. \ref{fig:n7782_ellipse}).

\begin{figure}    
\vspace*{7cm}    
\includegraphics{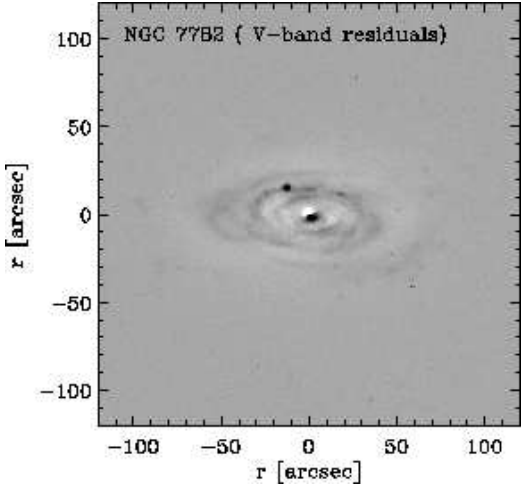}    
\caption[]{The residual image of NGC 7782     
  obtained by subtracting the model surface brightness to the observed
  one. The orientation of the image is the same of
  Fig. \protect\ref{fig:n7782_kinematics}.  The photometric parameters
  of the model are given in Table \protect\ref{tab:result2Dfit}.}
\label{fig:n7782_Vband_residuals}    
\end{figure}

The ellipticity increases from $0.18$ to $0.37$ in the first $10''$ 
($3.6$ kpc), dropping to $0.22$ further out.  It then increases again 
to $0.48$ and flattens around this value for $r \geq 30''$. The 
position angle is constant at $175^\circ$ at almost all radii, except 
for two abrupt changes at $6''$ and $18''$ by less than $20^\circ$. 
Isophotes are boxy-shaped around $6''$ and between $18''$ and $36''$, 
and almost elliptical elsewhere.  The surface-brightness profile is 
characterized by a plateau at $21.6$ \mas\ between $23''$ and 
$30''$. All these features can be explained as due to the 
spiral pattern (see Fig. \ref{fig:n7782_Vband_residuals}).  
 
The $V-$band surface-brightness radial profile of Koidaira et    
al. (1990) has been extracted along the galaxy major axis but it is not  
directly comparable with the ellipse-averaged one we have derived  
and shown in Fig. \ref{fig:n7782_ellipse}.    
      
The two-dimensional bulge-disc parametric decomposition of NGC 7782 
was performed adopting an exponential bulge since this yielded lower 
residuals.  The map of the residuals obtained as a difference between 
the observed and the model surface brightness is plotted in 
Fig. \ref{fig:n7782_Vband_residuals}. 
This figure shows the asymmetric 
pattern of the inner stellar spiral arms. 

\subsubsection{Ionized-gas distribution}    
\label{sec:n7782_Halpha}    
    
\begin{figure}    
\vspace*{14.5cm}    
\includegraphics{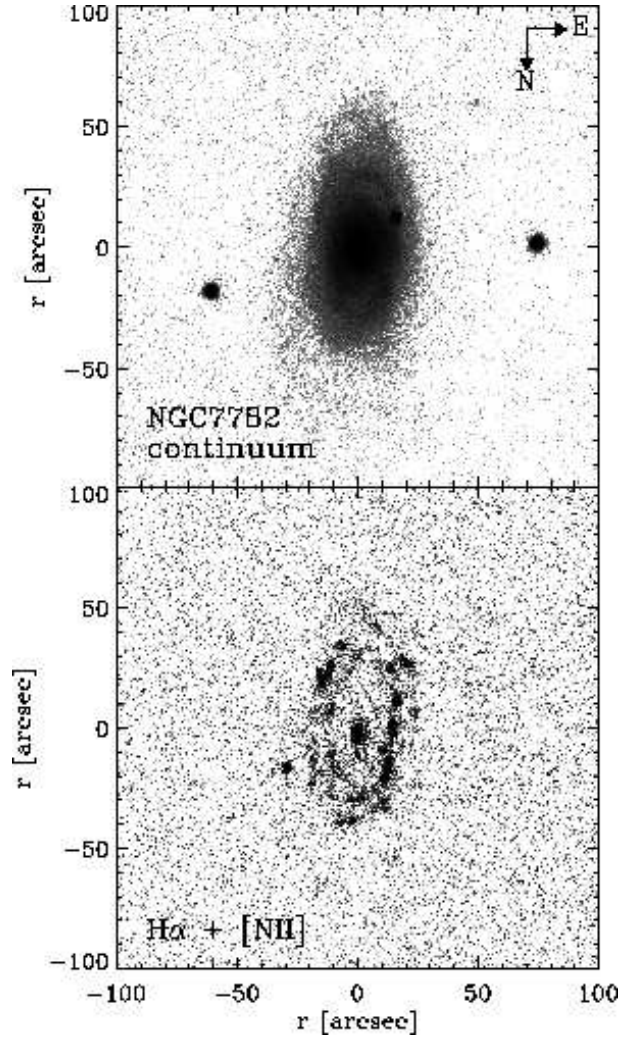}    
\caption[]{The stellar-continuum image ({\it upper panel\/}) and the    
  continuum-subtracted \ha$+$\nii\ emission image  ({\it lower panel\/})    
  of NGC 7782.}    
\label{fig:n7782_Halpha}    
\end{figure}

The ionized-gas distribution of NGC 7782 is shown in Fig. 
\ref{fig:n7782_Halpha}. The map of the \ha$+$\nii\ emission is 
characterized by the presence of a small and bright nuclear region 
aligned with the galaxy major axis, and two tightly wound spiral arms 
which can be followed for $360^\circ$ forming a sort of double 
ring-like structure. The gaseous arms are more symmetric than the 
stellar arms and do not extend to the inner region of the galaxy, starting 
outside the constant surface-brightness region at $30''$ from the 
centre.  
    
As in the case of NGC 772, the bright \ha$+$\nii\ nucleus of NGC 7782 
is characterized by a high ionized-gas velocity dispersion 
($\sigma_{\it gas}> 100$ \kms\ for $|r| \la 3''$) as shown in 
Fig. \ref{fig:n7782_kinematics}, suggesting the possible presence of 
pressure-supported gas in the bulge region.  In spite of its apparent 
clumpy distribution (Fig. \ref{fig:n7782_Halpha}) a diffuse component 
of ionized gas is also present in the disc of NGC 7782, as indicated 
by the fact that we are able to measure continuously the ionized-gas 
velocity parameter without gaps along the complete major axis out to 
$30''$ and $45''$ on the two sides of the nucleus 
(Fig. \ref{fig:n7782_kinematics}) respectively. The distribution of 
the \hii\ regions, following the continuum isophotes, indicates that 
the gas and stellar discs are coplanar. 
    
\subsubsection{Dynamical modelling}    
\label{sec:n7782_modelling}    
   
\begin{figure*}    
\vspace*{21.5cm}   
\includegraphics{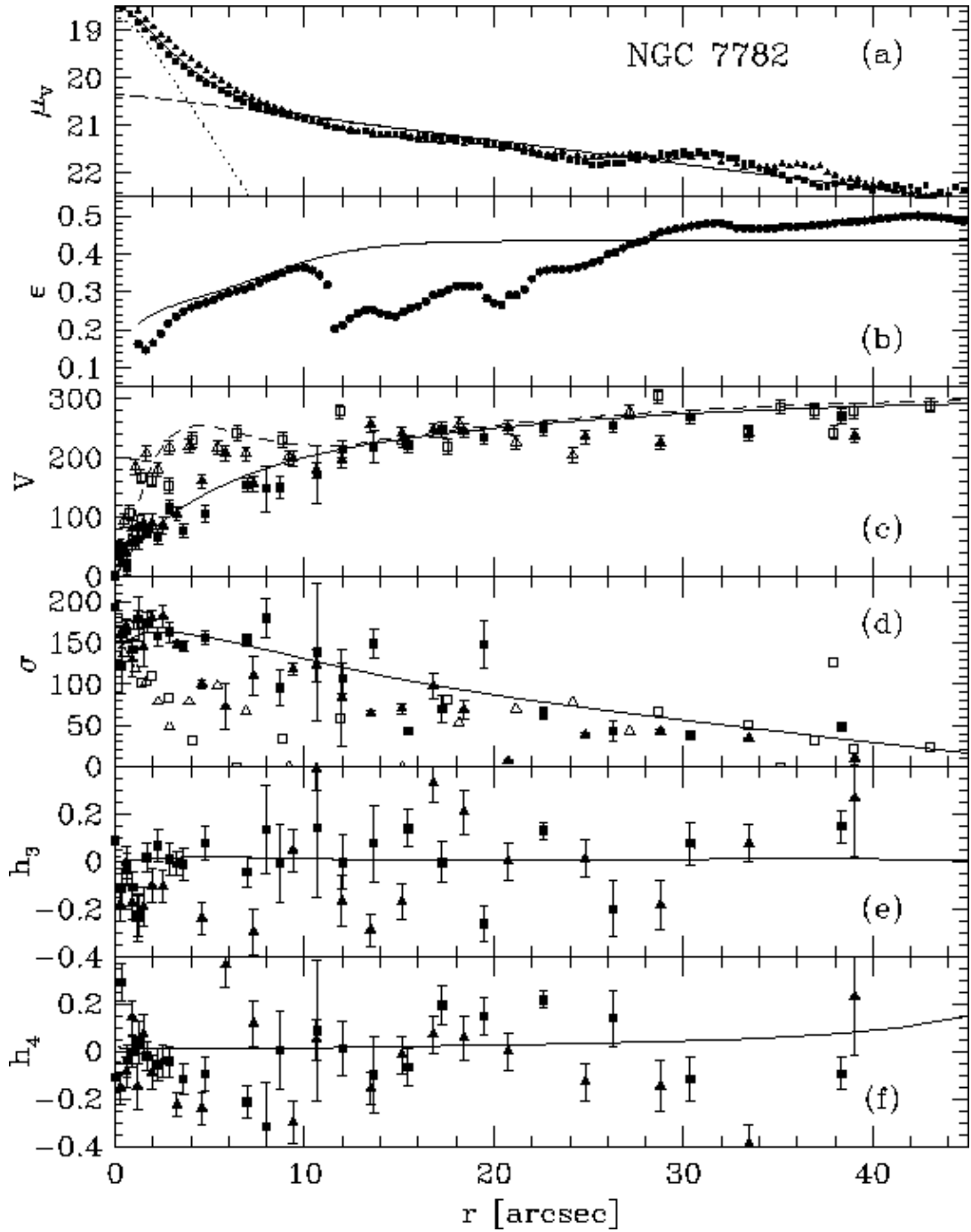}    
\caption[]{Same as Fig.~\protect\ref{fig:n772_model} but for NGC 7782.   
 The {\it squares\/} and {\it triangles\/} represent data measured on the    
 receding NW and approaching SE sides, respectively.}    
\label{fig:n7782_model}    
\end{figure*}

The comparison between the photometric and kinematic data and the 
best-fit photometric dynamical models we obtained for NGC 7782 is 
shown in Fig. \ref{fig:n7782_model}.  The mass-to-light ratios, the 
flattening and the masses of the bulge and disc stellar components 
derived using the dynamical modelling technique described in 
Sec. \ref{sec:dynmodels} are given in Table~\ref{tab:masses}. 
   
From both the kinematic and the photometric data shown in   
Fig.~\ref{fig:n7782_model} we can clearly distinguish   
the bulge and disc-dominated regions.    
 
An exponential and almost spherical bulge is the main photometric and 
dynamical component within $4''$ from the centre.  In this region the 
stellar velocity dispersion shows a plateau at 180 \kms\ and in 
agreement with the asymmetric drift effect calculated by the model the 
stellar rotation velocity is below the value of the circular 
velocity.  In the transition region bewteen $2'' \la |r| \la 7''$ the 
gas rotation curve shows small deviations ($|V_{\it gas}-V_{\it circ}|  
\la 50$ \kms) 
from pure circular rotation. Any evidence for possible non-circular 
motions is marginal and the increase of the gas velocity dispersion 
towards the nucleus can be better explained as due to a seeing effect. 
  
For $|r|>15''$ the surface brightness and ellipticity profiles of NGC 
7782 are dominated by the light contribution of the disc 
component. Therefore in the outer region asymmetric drift effect is 
negligible and both gas and stars appear to rotate close to the 
circular velocity. No dark matter halo is required to reproduced the 
observed velocity curve out to the last observed radius. 
  
Since the $h_3$ and $h_4$ values are poorly constrained by observations,  
it is not possible to draw any conclusion about 
anisotropies in the stellar velocity dispersion. 
 
\section{Discussion and conclusions}    
\label{sec:conclusions}    
 
Non-circular velocities of the ionized gas in the 
bulge-dominated region of early-to-intermediate type disc 
galaxies have been detected by different authors 
(Fillmore et al. 1986; Kent 1988; Kormendy \& Westpfahl 
1989). The discovery that in a large fraction of S0's the 
gas velocity dispersion remains as high as the stellar 
velocity dispersion over an extended radial range 
(Bertola et al. 1995) corroborated the early suggestions 
that at small radii gas could be dynamically hot rather 
than settled in a cold disc. In this framework, the 
dynamical modelling by Cinzano et al. (1999) showed that 
in the S0 galaxy NGC 4036 the difference between the 
circular velocity curve inferred from stellar kinematics 
and the rotation curve of the gas is substantially  
accounted for its high velocity dispersion. 
 
It is usually customary, in measuring mass distribution, 
to assume that the ionized gas is moving in circular 
orbits. It is therefore crucial to understand if and when 
the warm gas in galaxies has a rotation 
curve which rises more slowly than the gravitational 
equilibrium circular velocity. Ignoring this 
effect can lead to a significant underestimate of the masses 
of the bulge, disc and dark halo
(e.g. Kent 1988 for a 
discussion).  Up to now such a direct comparison between 
the circular velocity inferred from dynamical modelling 
based on stellar kinematics and surface photometry, and 
the observed gas rotation velocity has been done for only 
a small set of lenticulars and early-to-intermediate 
spirals (see Table \ref{tab:slowlyrising}). The three 
new cases discussed in this paper represent 
therefore a useful step in understanding how common the 
phenomenon of slowly-rising rotation curves is in 
bulge-dominated galaxies. 
 
For NGC 772, NGC 3898 and NGC 7782 we present a 
self-consistent Jeans model for the stellar kinematics, 
adopting the light distribution of bulge and disc derived 
by means of a two-dimensional parametric photometric 
decomposition of the observed $V-$band surface 
brightness.  This allows us not only to investigate the 
presence of non-circular gas motions, but also to derive 
the mass distribution of luminous and dark matter in 
these objects. 

\begin{table*}    
\caption{Early-to-intermediate type disc galaxies with available 
comparison 
between gas rotation velocity and circular speed}    
\begin{center}    
\begin{tabular}{lllrccrccccl}    
\noalign{\smallskip}    
\hline    
\noalign{\smallskip}    
\multicolumn{1}{c}{Object} &     
\multicolumn{2}{c}{Morphological Type} &    
\multicolumn{1}{c}{$D$} &   
\multicolumn{1}{c}{$M^0_{B_T}$}  & 
\multicolumn{1}{c}{$M_{\it bulge}/M_{\it disc}$}  &   
\multicolumn{1}{c}{$r_e$} &    
\multicolumn{1}{c}{$r_\mu$} &  
\multicolumn{1}{c}{$V_{\it gas}<V_{\it circ}$} &  
\multicolumn{1}{c}{$r_V$} &   
\multicolumn{1}{c}{$r_\sigma$} & 
\multicolumn{1}{c}{Ref.} \\ 
\noalign{\smallskip}    
\multicolumn{1}{c}{} &    
\multicolumn{1}{c}{[RSA]} &    
\multicolumn{1}{c}{[RC3]} &    
\multicolumn{1}{c}{[Mpc]} &    
\multicolumn{1}{c}{[mag]}&  
\multicolumn{1}{c}{} &  
\multicolumn{1}{c}{[kpc]} & 
\multicolumn{1}{c}{[kpc]} & 
\multicolumn{1}{c}{} & 
\multicolumn{1}{c}{[kpc]} & 
\multicolumn{1}{c}{[kpc]} &  
\multicolumn{1}{c}{} \\ 
\noalign{\smallskip}    
\multicolumn{1}{c}{(1)} &    
\multicolumn{1}{c}{(2)} &    
\multicolumn{1}{c}{(3)} &    
\multicolumn{1}{c}{(4)} &    
\multicolumn{1}{c}{(5)} &    
\multicolumn{1}{c}{(6)} &    
\multicolumn{1}{c}{(7)} &    
\multicolumn{1}{c}{(8)} &    
\multicolumn{1}{c}{(9)} &    
\multicolumn{1}{c}{(10)} & 
\multicolumn{1}{c}{(11)} & 
\multicolumn{1}{c}{(12)} \\   
\noalign{\smallskip}    
\hline    
\noalign{\smallskip}    
NGC~772  & Sb(rs)I      & .SAS3..   & 34.7 & -21.61 & 2.11 & 11.0 & 6.2 
& no  & -      & 0.5 &1, $V$\\  
NGC~2179 & Sa           & .SAS0..   & 33.7 & -19.81 & 2.80 &  2.0 & 5.7 
& no  & -      & 0.2 &2, $R$\\  
NGC~2775 & Sa(r)        & .SAR2..   & 13.9 & -19.91 & 1.39 &  3.6 & 4.4 
& yes & 2.0    & -   &2, $r$\\  
NGC~3898 & SaI          & .SAS2..   & 17.1 & -19.84 & 3.55 &  1.6 & 2.3 
& yes & 0.7    & -   &1, $V$\\  
NGC~4036 & S0$_3$(8)/Sa & .L..$-$.. & 20.1 & -20.03 & 2.04 &  1.2 & 0.7 
& yes & 1.0    & 0.4 &3, $V$\\  
NGC~4450 & Sab pec      & .SAS2.    & 17.0 & -20.40 & 2.69 &  1.7 & 1.4 
& yes & $>5.0$ & ?   &4, $B$\\  
NGC~4569 & Sab(s)I-II   & .SXT2..   & 17.0 & -21.36 & 0.08 &  0.3 & 0.4 
& yes & 2.6    & ?   &4, $B$\\  
NGC~5055 & Sbc(s)II-III & .SAT4..   &  7.5 & -20.35 & 0.12 &  0.4 & 0.3 
& yes & 0.5    & ?   &4, $B$\\  
NGC~7782 & Sb(s)I-II    & .SAS3..   & 74.8 & -21.95 & 0.24 &  1.1 & 1.5 
& no  & -      & 0.4 &1, $V$\\  
\noalign{\smallskip}    
\hline    
\noalign{\smallskip}    
\noalign{\smallskip}    
\label{tab:slowlyrising}    
\end{tabular}    
\begin{minipage}{16.cm}    
NOTE -- col. 2: morphological type from RSA;  
col. 3: morphological type from RC3;  
col. 4: distance derived as $V_0/H_0$ with $V_0$ the velocity relative 
 to the centroid of the Local Group obtained from the heliocentric 
 systemic velocity as in RSA and $H_0 = 75$ \kms\ Mpc$^{-1}$. The  
 heliocentric velocity of NGC 5055 has been taken from RC3. 
 For NGC 4450 and NGC 4569 which belong to the Virgo Cluster the 
 distance has been taken following Freedman et al. (1994); 
col. 5: absolute corrected $B$ magnitude from $B^0_T$ in RC3; 
col. 6: bulge-to-disc mass ratio; 
col. 7: effective radius of the $r^{1/4}$ bulge. For NGC 7782 
 the bulge has an exponential surface-brightness profile; 
col. 8: radius at which $\mu_{\it bulge} = \mu_{\it disc}$; 
col. 9: yes = the ionized-gas component shows a slowly-rising 
 rotation curve (i.e. $V_{\it gas} < V_{\it circ}$ at small radii),  
 no = the gas rotation traces circular speed at all 
 radii ($V_{\it gas} = V_{\it circ}$ at all radii); 
col. 10: radial range in which $V_{\it gas} < V_{\it circ}$; 
col. 11: radial range in which $\sigma_{\it gas} \simeq \sigma_{\it 
stars}$, 
 ? = the radial profile of the ionized-gas velocity dispersion is not 
 available; 
col. 12: references and bands of the data [1 = this paper, 2 = Corsini 
 et al. (1999), 3 = Cinzano et al. (1999), 4 = Fillmore et al. (1986)]. 
\end{minipage}    
\end{center}    
\end{table*}

In NGC~772 the observed velocity dispersion of the 
ionized gas is comparable to the stellar velocity dispersion  
and is far higher than expected from the thermal 
motions or small-scale turbulence, typically observed in 
a gaseous disc. However, dynamical modelling showed that the 
rotation of the ionized-gas traces the circular velocity 
and the central rise observed in its velocity dispersion 
is due to seeing effects, which serves as a warning against 
over interpretation. 
 
On the contrary, the gas rotation curve of NGC 3898 can 
be classified as `slowly-rising', since for the inner 
$8''$ ionized gas is rotating more slowly than the circular 
velocity predicted by dynamical modelling.  NGC 3898 is 
the only galaxy of our sample for which previous 
determinations of the masses of the bulge and disc 
components are available in the literature (Kent 1988; 
Fillmore et al. 1986; Moriondo et al. 1998b).  
Comparison shows that NGC 3898 is a clearcut example 
that mass decomposition based only on emission-line 
rotation curves can be unreliable, at least for 
bulge-dominated galaxies. This error was made  
and discussed by Kent 
(1988) and Moriondo et al. (1998b), who assumed the gas 
to be rotating at circular velocity at all radii and 
found extraordinarily low values for the mass-to-light 
ratio of the spheroidal component, as if NGC 3898 was an 
almost `bulge-less' galaxy in spite of its overall 
morphology (see Table \ref{tab:properties} and 
Figs. \ref{fig:n3898_kinematics},  
and \ref{fig:n3898_Halpha}) 
and bulge-to-disc luminosity ratio ($B/D=2$).  The 
asymmetry of the gas rotation curve of NGC 3898 (with a 
maximum $\Delta V$ between the two sides of about 80 
\kms\ at $6''$ from the centre), and the strong 
difference between the gas ($\sigma_{\it gas} \leq 90$ 
\kms) and the stellar velocity dispersion ($\sigma_{\it 
stars} > 120$ \kms\ with a central peak to $210$ \kms) in 
the innermost $5''$ (where we observe almost the same 
velocity gradient for both gas and stars) are an 
indication that along with random motions other 
phenomena could also be contributing to the slowly rise of the gas 
velocity. This is the case of other galaxies with 
slowly-rising rotation curves, such as the Sa NGC 2775 
(Corsini et al. 1999) and the S0 NGC 4036 (Cinzano et 
al. 1999). The asymmetry of the inner parts of the gas 
rotation curve of NGC 2775 has been explained as due to a 
gaseous component which is not rotating in the galaxy 
plane, and the possible presence of drag forces between 
the ionized gas and the hot component of the interstellar 
medium of NGC 4036 has been suggested (for a discussion 
of the phenomenon see Cinzano et al. 1999). 
 
Finally, in NGC~7782 the gas rotation curve shows small deviations 
from pure circular rotation only in the transition region 
between bulge and disc.  Any evidence for possible 
non-circular motions is marginal and the sharp increase 
of the gas velocity dispersion towards the nucleus can be 
better explained as due to a seeing effect as in the case 
of NGC 772. 
 
As far as the presence of dark matter is concerned, we 
infer that the mass is essentially traced by light in 
NGC 772 and NGC 7782, where gas rotation velocities were 
observed out to $0.2\,R_{25}$ and $0.6\,R_{25}$, 
respectively. For NGC 3898 the combined stellar and 
gaseous rotation data (which extend out to $0.2\,R_{25}$ 
and $0.7\,R_{25}$ respectively) require the presence of a 
massive dark halo. This result is based on the idea that 
gas kinematics at large radii is representative of a 
dynamically cold disc supported by rotation, as suggested 
by the high gas rotation velocity and low velocity 
dispersion ($V/\sigma\simeq10$) for $r>30''$ 
(corresponding to about $0.2\,R_{25}$). This result also 
qualitatively agrees with the general dark matter 
scenario, with the less massive galaxies being the more 
dark-matter dominated (Salucci \& Persic 1999), although 
these data do not permit a careful measurement of the 
dark halo properties. 
 
Kent (1988) found that, out of the 14 Sa galaxies he 
modeled, six show gas rotation curves that rise too 
slowly for a constant mass-to-light ratio implying they are 
respectively too low for the bulges and too high for the 
disc components. According to 
Table~\ref{tab:slowlyrising}, the gas motion is not circular 
in the inner regions of 6 of the 9 
early-to-intermediate disc galaxies, for which dynamical 
modelling allows the direct comparison between the gaseous 
and circular speeds (among these only NGC 3898 belongs 
also to Kent's sample).

Without pretending to draw any statistical conclusion 
based on so small number of objects, it nevertheless 
seems probable that non-circular gas velocities could be a 
common feature in the central kiloparsec of disc galaxies 
ranging from S0's to Sb's. However, according to the 
available data, it is not possible to derive a 
straightforward correlation with the galaxy morphological 
type, total luminosity, bulge-to-disc mass ratio or bulge 
size. It seems that even if the gas may be supported by 
non-circular random motions, this type of dynamical-pressure 
support may not be the only effect for the 
`slowly-rising' gas rotation curves. In fact, the central 
gas velocity dispersion is too low to account for the 
difference between gas and circular speeds in two of the 
three galaxies (namely NGC 2775, and NGC 3898) shown in 
Table \ref{tab:slowlyrising} for which the gas 
velocity falls short of the circular speed and the 
velocity-dispersion profile of the gaseous component is available. 
Dynamical modelling based both on stellar and gaseous 
kinematics of a larger number of bulge-dominated galaxies is 
needed to understand the links between slowly-rotating 
gas and bulge properties, as a first step to put 
constraints on its physical nature (e.g. Mathews 1990) 
and origin (e.g. Bertola et al. 1995). 
 
\medskip 
\noindent   
{\bf Acknowledgements.}  
 
\noindent   
We are grateful to Prof. P.A. Strittmatter, Director of
the Steward Observatory and to Dr. G.V. Coyne, S.J.,
Director of the Vatican Observatory for the allocation of
time for our observations. WWZ acknowledges the support
of the grant 7914 of the {\sl Jubil\"aumsfonds der
Oesterreichischen Nationalbank}. This work was partially
supported by grant PB97-0214 of the Spanish DGES.

\end{document}